\documentclass{article}
\usepackage[a4paper,margin=1in,footskip=0.25in]{geometry}
\usepackage{graphicx} % Required for inserting images
\usepackage{amsmath}
\usepackage{authblk}
\usepackage{amssymb}
\usepackage{mathtools}
\usepackage{stmaryrd}
\usepackage{hyperref}
\usepackage{xfrac}
\usepackage{natbib}
\usepackage{siunitx}
\usepackage{color}
\usepackage{tcolorbox}
\usepackage{subcaption}
\usepackage{tikz}
\usepackage{multirow}
\usepackage{tikz-3dplot}
 \usetikzlibrary{calc}
 \usetikzlibrary{patterns}

\def\correspondingauthor{\footnote{Corresponding author: louis.libat2@univ-eiffel.fr}}

\title{A space-time extension of a conservative two-fluid cut-cell method for moving diffusion problems}
\author[1]{Louis Libat \correspondingauthor}
\author[1]{Can Selçuk}
\author[1]{Eric Chénier}
\author[1]{Vincent Le Chenadec}
\affil[1]{MSME, Université Gustave Eiffel, UMR CNRS 8208, Marne-la-Vallée, 77454, France}

\date{\today}

\begin{document}

\maketitle

\begin{abstract}
We present a space-time extension of a conservative Cartesian cut-cell
finite-volume method for two-phase diffusion problems with prescribed interface motion.
The formulation follows a two-fluid approach: one scalar field is solved in
each phase with discontinuous material properties, coupled by sharp interface
conditions enforcing flux continuity and jump laws.
To handle moving boundaries on a fixed Cartesian grid, the discrete balance is
written over phase-restricted space-time control volumes, whose geometric
moments (swept volumes and apertures) are used as weights in the finite-volume
operators. This construction naturally accounts for the creation and
destruction of cut cells (fresh/dead-cell events) and yields strict discrete
conservation. The resulting scheme retains the
algebraic structure of the static cut-cell formulation while incorporating
motion through local geometric weights and interface coupling operators.
A series of verification and validation tests in two and three dimensions
demonstrate super-linear accuracy in space, robust behavior under repeated
topology changes and conservation across strong coefficient jumps and moving
interfaces. The proposed space-time cut-cell framework provides a conservative
building block for multiphase transport in evolving geometries and a foundation
for future free-boundary extensions such as Stefan-type phase change.
\end{abstract}

\section{Introduction}
\label{sec:introduction}

Two-phase scalar transport with sharp interface conditions in temperature or
concentration underpins a wide range of natural and industrial processes, from
solidification and melting in metallurgical applications to reactive separations
and dissolution problems in chemical engineering \cite{bird_transport_2007}.
In such settings, two media with distinct material properties (e.g.\ diffusivity,
capacity, solubility) are separated by a moving boundary across which the scalar
field may remain continuous or may satisfy partition laws, while the diffusive
flux obeys a conservation constraint.
Predictive simulation therefore hinges on the ability to enforce these jump and
continuity relations sharply, while preserving the local balance of the
transported quantity.

A central numerical difficulty is geometric: the relevant physics often
concentrates in the vicinity of material boundaries, where large gradients and
boundary fluxes control the global exchange rates.
This is well known in boundary-dominated flow and transport configurations,
for instance when viscous stresses and pressure drag are set by near-wall
dynamics or when conjugate heat transfer near solid surfaces controls thermal
loads in engineering devices.
In diffusion-dominated multiphase problems, the same observation holds: accuracy
and robustness are largely determined by how the method represents curved
boundaries and how it enforces interfacial fluxes.

Body-fitted discretizations can represent boundaries accurately but they
typically require mesh motion and/or remeshing when interfaces deform or
translate, which complicates robustness, conservation and automation.
Unstructured techniques can conform to complex geometries and are very effective
for general shapes, including (piecewise) curved surfaces \cite{hughes_isogeometric_2005},
but high-quality mesh generation remains challenging and time-consuming and the
explicit storage of element connectivity increases both algorithmic complexity
and computational cost \cite{mavriplis_unstructured_1995}.
These limitations are one of two compelling arguments for structured Cartesian
meshes, the second being the simplicity and efficiency of many solvers and data
structures on such grids.

This has motivated fixed-grid approaches where the interface is embedded in a
Cartesian background mesh.
On non-conforming grids, immersed-boundary and ghost-cell techniques enforce
boundary and interface constraints by modifying stencils or introducing forcing
terms \cite{mittal_immersed_2005,peskin_flow_1972,gabbard_high-order_2024,gibou_second-order-accurate_2002}.
Diffuse-interface variants were historically popular in immersed-boundary
formulations \cite{peskin_flow_1972} and have been generalized through fictitious
domain and Lagrange-multiplier strategies \cite{glowinski_fictitious_1994,taira_immersed_2007}.
While these approaches are attractive for their flexibility, they may compromise
strict local conservation or introduce small but systematic errors near the
interface when strong coefficient jumps are present.

Embedded-boundary (cut-cell) finite-volume methods provide an alternative that
is particularly well suited to diffusion-dominated transport.
In cut-cell approaches, the physical control volumes are obtained by intersecting
Cartesian cells with the true geometry and the discrete equations are derived
directly from integral balances so that conservation is enforced by construction.
Pioneering Cartesian embedded-boundary formulations for elliptic operators were
proposed by Johansen and Colella \cite{johansen_cartesian_1998} and extended to
finite-volume advection-diffusion in irregular domains by Calhoun
et al.\ \cite{calhoun_cartesian_2000}.
For diffusion and Poisson problems, conservative cut-cell schemes and accurate
boundary-flux reconstructions have been developed and analyzed in a variety of
configurations \cite{mccorquodale_cartesian_2001,schwartz_cartesian_2006,mccorquodale_high-order_2011}.
These results make cut-cell discretizations an appealing foundation for
two-phase scalar transport: they naturally incorporate discontinuous material
properties, enforce boundary conditions through physical fluxes on embedded
faces and recover standard second-order formulas away from the boundary while
preserving local conservation in strongly irregular cut cells.
Moreover, the underlying two-fluid Cartesian cut-cell framework has already been
developed and validated \cite{libat_cartesian_2025} for fixed-boundary diffusion problems,
including sharp interface coupling and conservative flux enforcement on embedded
boundaries.

The situation becomes substantially more delicate when the interface moves.
As the geometry evolves, the cut-cell volumes and face apertures vary in time,
and cells can suddenly appear or disappear as the boundary sweeps through the
mesh (the fresh and dead cut-cell events \cite{farhat_discrete_2001}).
Without a consistent treatment of these geometric changes, purely spatial
formulations can exhibit spurious gain or loss of scalar mass/energy.
This issue is closely related to the geometric conservation law on moving meshes
\cite{farhat_discrete_2001} and has been addressed in several moving-boundary
and fluid-structure contexts \cite{ho_discrete_2021}.
From the viewpoint of conservation, a natural remedy is to formulate the method
directly in space-time: balances are written over swept control volumes on each
time slab, so that changes in volume and aperture are accounted for by
construction through a discrete Reynolds transport theorem and an exact
geometric conservation law.

Following Tarzia's terminology \cite{tarzia_bibliography_2000}, two-phase
diffusion problems can be classified into three categories:
(i)~\emph{fixed-boundary problems}, where the geometry is time-independent and
only classical boundary conditions (Dirichlet/Neumann/Robin) or jumps law on static material
interfaces must be enforced;
(ii)~\emph{moving-boundary problems}, where the interface motion is prescribed
(e.g.\ translating, oscillating or rigid-body kinematics), which already
induces repeated cut-cell topology changes;
(iii)~\emph{free-boundary problems}, where the interface position is part of the
solution, as in Stefan-type phase change driven by a latent-heat balance
\cite{stefan_ueber_1891,alexiades_mathematical_2018}.
The present paper targets the intermediate but essential setting (ii): prescribed
motion of a sharp interface, viewed as a controlled environment to design and
validate conservative space-time operators that will also be required for
future free-boundary (Stefan) extensions.

We present a space-time extension of a conservative two-fluid cut-cell
diffusion method for moving geometries.
The guiding principle is to retain the simplicity and efficiency of a Cartesian
background mesh, while representing the moving interface sharply through
phase-restricted control volumes and enforcing coupling through local jump
operators.
Geometrically, the method relies on a finite set of robust moments (volumes,
apertures, centroids and staggered moments as needed) computed from intersection
operations; these geometric fields are the only inputs required to modify the
standard finite-volume formulas in the vicinity of the moving boundary and the
operators degenerate to their classical mesh-aligned counterparts when the
interface aligns with grid faces.

The main contributions are:
\begin{itemize}
\item A space-time finite-volume formulation on phase-restricted control
volumes, in which the scalar balance is enforced over swept space-time regions
so that geometric changes (fresh/dead cut cells) are handled conservatively by
construction.

\item A two-fluid interface treatment that preserves sharp jump/continuity
conditions using local coupling operators supported by cut-cell geometric
weights, allowing discontinuous diffusivities and interfacial transfer laws to
be enforced without smearing.

\item A practical implementation strategy based on robust geometric moments and
intersection operations, suitable for repeated prescribed interface motion on a
Cartesian mesh.
\end{itemize}

\medskip
The remainder of the paper is organized as follows.
Section~\ref{sec:continuum} introduces the two-fluid diffusion model, the jump and
boundary conditions and the prescribed kinematics of the moving interface.
Sections~\ref{sec:discrete} and~\ref{sec:moving-cutcell} detail the construction of
phase-restricted geometric moments and discrete operators and derive the
space-time cut-cell finite-volume discretization.
Section~\ref{sec:nummotion} presents numerical experiments and convergence studies.
Conclusions and perspectives are provided in Section~\ref{sec:conclusion}.

\section{Continuum modeling}
\label{sec:continuum}

\begin{figure}
\centering
\includegraphics[width=0.5\linewidth]{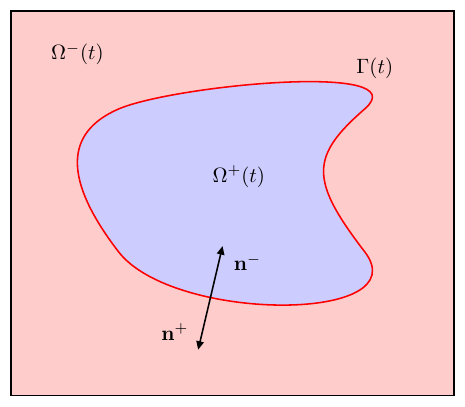}
\caption{Fixed background domain $\Omega$ with a prescribed moving interface
$\Gamma(t)$ partitioning $\Omega$ into $\Omega^-(t)$ and $\Omega^+(t)$. Normals
$\mathbf n^\pm$ point outward of their respective phases.}
\label{fig:open-sine-interface}
\end{figure}

We consider diffusive transfers between two immiscible phases in a fixed
background domain $\Omega \subset \mathbb{R}^d$ ($d\in\{1,2,3\}$). The geometry
evolves in time through a prescribed sharp interface $\Gamma(t)$ that partitions
$\Omega$ into two time-dependent subdomains occupied by the red and blue
phases, denoted $\Omega^-(t)$ and $\Omega^+(t)$, respectively (Figure~\ref{fig:open-sine-interface}).
The immiscibility and saturation conditions read
\begin{equation}
\Omega^-(t)\cap\Omega^+(t)=\emptyset,
\label{eq:immiscibility}
\end{equation}
\begin{equation}
\Omega^-(t)\cup\Gamma(t)\cup\Omega^+(t)=\Omega.
\label{eq:saturation}
\end{equation}
Superscripts ``$-$'' and ``$+$'' denote phase-wise quantities and
$\mathbf n^\pm$ are the unit normals on $\Gamma(t)$ pointing outward of
$\Omega^\pm(t)$. We take the red phase as reference and define
\(
\mathbf n := \mathbf n^-  \text{ on }\Gamma(t).
\)
The interface motion is prescribed through an interfacial velocity
$\mathbf w(t,\mathbf x)$, with normal component
\begin{equation}
w(t,\mathbf x) := \mathbf w(t,\mathbf x)\cdot \mathbf n(t,\mathbf x),
\qquad \mathbf x\in \Gamma(t).
\label{eq:normal-velocity}
\end{equation}
In this work, $\Gamma(t)$ (and thus $w$) is given and the transport problem is
solved on the evolving domains $\Omega^\pm(t)$.

\subsection{Bulk transport equations}

Let $\phi$ denote a generic scalar (temperature $T$ or concentration $c$). In
each phase, $\phi^\pm$ satisfies the diffusion balance
\begin{equation}
C^\pm \,\frac{\partial \phi^\pm}{\partial t} + \nabla\cdot \mathbf q^\pm = r^\pm,
\qquad t>0,\ \mathbf x\in\Omega^\pm(t),
\label{eq:bulk-balance}
\end{equation}
closed by a purely diffusive constitutive law (Fourier/Fick),
\begin{equation}
\mathbf q^\pm := -K^\pm \nabla \phi^\pm.
\label{eq:constitutive}
\end{equation}
Here $C^\pm$ denotes the capacity (e.g.\ $\rho^\pm c_p^\pm$ for heat, $\rho^\pm$
for species) and $K^\pm$ the mobility (e.g.\ $k^\pm$ or $D^\pm$). In the present
study, $C^\pm$ and $K^\pm$ are taken constant within each phase to retain a
linear bulk operator; $r^\pm$ denotes a volumetric source term. Initial
conditions are prescribed at $t=0$ on $\Omega^\pm(0)$.

\subsection{Interfacial conditions}
\label{sec:interfacial-conditions}

We recall the definition of the $\lambda$-weighted jump at an
interfacial point $\mathbf x\in\Gamma(t)$ by
\begin{equation}
\llbracket \phi \rrbracket^\lambda(t,\mathbf x)
:=
\lim_{\varepsilon\to 0^+}
\Bigl[
\phi^+\bigl(t,\mathbf x-\varepsilon \mathbf n^+(t,\mathbf x)\bigr)
-
\bigl(\lambda\,\phi^-\bigr)\bigl(t,\mathbf x-\varepsilon \mathbf n^-(t,\mathbf x)\bigr)
\Bigr].
\label{eq:weighted-jump}
\end{equation}
When $\lambda\equiv 1$, we recover the standard jump notation
$\llbracket \phi \rrbracket := \llbracket \phi \rrbracket^1$.

Conservation at the moving interface (neglecting interfacial storage, line
fluxes and interfacial sources) yields the continuity of the normal flux in
the interface frame. With $w$ defined by \eqref{eq:normal-velocity}, this reads
\begin{equation}
\llbracket \mathbf q\cdot \mathbf n - C\,\phi\,w \rrbracket(t,\mathbf x) = 0,
\qquad t>0,\ \mathbf x\in\Gamma(t).
\label{eq:surface-balance}
\end{equation}
The term $C\,\phi\,w$ accounts for the transport of $\phi$ induced by the motion
of the interface and is the key contribution distinguishing the moving-geometry
setting from the static one.

Equation~\eqref{eq:surface-balance} must be complemented by an additional
interfacial law to close \eqref{eq:bulk-balance}. Since $\Gamma(t)$ is
prescribed here, a single scalar relation is sufficient. For conjugate heat
transfer one typically enforces continuity (possibly with a prescribed jump
$f$),
\begin{equation}
\llbracket \phi \rrbracket(t,\mathbf x) = f(t,\mathbf x),
\qquad t>0,\ \mathbf x\in\Gamma(t),
\label{eq:continuity}
\end{equation}
while for conjugate mass transfer one may use Henry's law,
\begin{equation}
\llbracket \phi \rrbracket^{\mathrm{He}}(t,\mathbf x) = f(t,\mathbf x),
\qquad t>0,\ \mathbf x\in\Gamma(t),
\label{eq:henry}
\end{equation}
where $\lambda=\mathrm{He}$ is Henry's coefficient (with $\mathrm{He}\to 1$ when
solubilities match). The right-hand side $f$ allows for interfacial jumps
associated with thin-layer effects or effective interfacial resistances.

Finally, we emphasize that the present work follows a two-fluid formulation:
the bulk equations are solved separately in $\Omega^\pm(t)$ and coupled through
the sharp conditions \eqref{eq:surface-balance}-\eqref{eq:henry}. This
preserves phase-wise modeling flexibility and avoids the introduction of
effective properties in interfacial control volumes.

\section{Discrete geometric representation}
\label{sec:discrete}

The space-time formulation of the continuum model developed in Section~\ref{sec:continuum} is discretized on a fixed Cartesian background grid, while the physical geometry is represented
by a prescribed time-dependent interface $\Gamma(t)$. The resulting phase domains
$\Omega^\pm(t)$ therefore evolve in time and continuously create and remove phase-restricted control volumes. Since the static geometric constructions, interface representations and
moment-based operator assembly are described in detail in \cite{libat_cartesian_2025},
we recall here only the minimal geometric notation required for the moving case
and for the definition of space-time cut-cell quantities used in
Section~\ref{sec:moving-cutcell}.

\subsection{Geometrical foundations (recall)}

For clarity, we present the two-dimensional setting ($d=2$), the extension to
$d=3$ is feasible. Let
$X=(x_{1/2},x_{3/2},\ldots,x_{N+1/2})$ and
$Y=(y_{1/2},y_{3/2},\ldots,y_{M+1/2})$ be strictly increasing grid coordinates.
The Cartesian cells are
\[
\Omega_{i,j}=\Delta_i^1\times\Delta_j^2,
\qquad
\Delta_i^1=(x_{i-1/2},x_{i+1/2}),\quad
\Delta_j^2=(y_{j-1/2},y_{j+1/2}),
\]
for $(i,j)\in[1\ldots N]\times[1\ldots M]$. At any time $t$, the phase-restricted
cells and the local interface segment are defined by
\begin{equation}
\Omega_{i,j}^\pm(t)=\Omega_{i,j}\cap \Omega^\pm(t),
\qquad
\Gamma_{i,j}(t)=\Omega_{i,j}\cap \Gamma(t),
\label{eq:cutcell_moving}
\end{equation}
so that (by construction) the disjoint decomposition holds:
\begin{equation}
\Omega_{i,j}=\Omega_{i,j}^-(t)\ \cup\ \Gamma_{i,j}(t)\ \cup\ \Omega_{i,j}^+(t).
\label{eq:cell_partition_moving}
\end{equation}
A cell is said to be \emph{pure} in phase $\pm$ at time $t$ if
$\Omega_{i,j}^\pm(t)=\Omega_{i,j}$ and \emph{mixed} if $\Gamma_{i,j}(t)\neq\emptyset$.
In the moving setting, these labels may change with time.

Intersections of Cartesian faces with $\Omega^\pm(t)$ are denoted similarly.
For example, the phase-restricted vertical and horizontal faces are
\[
\Sigma^{1\pm}_{i-1/2,j}(t)
=
\bigl(\{x_{i-1/2}\}\times \Delta_j^2\bigr)\cap \Omega^\pm(t),
\qquad
\Sigma^{2\pm}_{i,j-1/2}(t)
=
\bigl(\Delta_i^1\times \{y_{j-1/2}\}\bigr)\cap \Omega^\pm(t),
\]
with analogous definitions for the opposite faces. 

These subsets induce the
standard partition of the phase boundary (vertices excluded),
\begin{equation}
\partial \Omega_{i,j}^\pm(t)\ =
\Gamma_{i,j}(t)\ \cup\
\Sigma^{1\pm}_{i-1/2,j}(t)\ \cup\ \Sigma^{1\pm}_{i+1/2,j}(t)\ \cup\
\Sigma^{2\pm}_{i,j-1/2}(t)\ \cup\ \Sigma^{2\pm}_{i,j+1/2}(t),
\label{eq:boundary_partition_moving}
\end{equation}
which is the starting point for the discrete flux balances.

We also recall 
\[
\Sigma^{1\pm}_j(t,x) := \bigl(\{x\}\times\Delta_j^2\bigr)\cap\Omega^\pm(t),
\qquad
\Sigma^{2\pm}_i(t,y) := \bigl(\Delta_i^1\times\{y\}\bigr)\cap\Omega^\pm(t),
\]
\[
\Omega^{1\pm}_j\bigl(t,\Delta\bigr) := \bigl(\Delta\times\Delta_j^2\bigr)\cap\Omega^\pm(t),
\qquad
\Omega^{2\pm}_i\bigl(t,\Delta\bigr) := \bigl(\Delta_i^1\times\Delta\bigr)\cap\Omega^\pm(t).
\]

Finally, we introduce a strictly increasing sequence of time instants
\[
0=t_0 < t_1 < \cdots < t_n < \cdots,
\]
and denote $\Delta t_n := t_{n+1}-t_n$.

\begin{figure}[!t]
\centering
\includegraphics[width=0.6\textwidth]{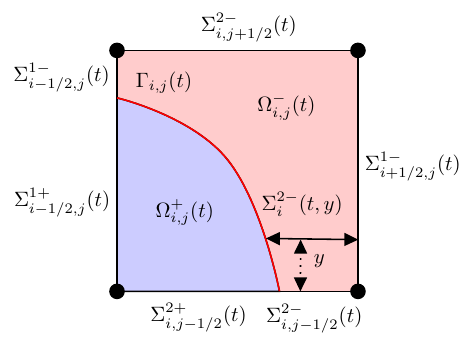}
\caption{Geometry definitions for a moving embedded interface on a Cartesian cell. \eqref{eq:boundary_partition_moving}.}
\label{fig:geom_foundations_cutcell}
\end{figure}

\subsection{Reduced geometric description}
\label{sec:moments}

As in the companion static paper, all discrete operators are expressed using a
reduced set of geometric moments. For any measurable set
$\Xi\subset\mathbb{R}^d$ and any integrable function $f$ (typically a monomial),
we use the shorthand notation
\begin{equation}
\left\langle \Xi, f \right\rangle
:= \int_{\mathbf x\in \Xi} f(\mathbf x)\,d\mathbf x .
\label{eq:integral}
\end{equation}
This notation is employed to define phase-restricted
cell volumes, face apertures and first moments over the time-dependent sets
$\Omega^\pm_{i,j}(t)$ and $\Sigma^{k\pm}(t)$, from which centroids and auxiliary
moments are derived.

For completeness, we recall the spatial low-order moments used by the static
operators (see the companion paper for the full discussion and computation).
On the Cartesian cell $\Omega_{i,j}$, the cell volume is
\[
V_{i,j} := \langle \Omega_{i,j},1\rangle
= (x_{i+\frac12}-x_{i-\frac12})(y_{j+\frac12}-y_{j-\frac12}),
\]
and the phase-restricted volumes (time-dependent in the moving setting) are
\begin{equation}
V^\pm_{i,j}(t) := \langle \Omega^\pm_{i,j}(t),1\rangle .
\label{eq:volume_cut}
\end{equation}
The cell center coordinates are
\[
x_i := \frac{\langle \Omega_{i,j},x\rangle}{\langle \Omega_{i,j},1\rangle}
= \frac{x_{i-\frac12}+x_{i+\frac12}}{2},
\qquad
y_j := \frac{\langle \Omega_{i,j},y\rangle}{\langle \Omega_{i,j},1\rangle}
= \frac{y_{j-\frac12}+y_{j+\frac12}}{2},
\]
and the phase centroids are defined by
\begin{equation}
x^\pm_{i,j}(t) :=
\begin{cases}
\langle \Omega^\pm_{i,j}(t),x\rangle / V^\pm_{i,j}(t),
& \text{if }V^\pm_{i,j}(t)\neq 0,\\
x_i, & \text{otherwise},
\end{cases}
\qquad
y^\pm_{i,j}(t) :=
\begin{cases}
\langle \Omega^\pm_{i,j}(t),y\rangle / V^\pm_{i,j}(t),
& \text{if }V^\pm_{i,j}(t)\neq 0,\\
y_j, & \text{otherwise}.
\end{cases}
\label{eq:bulk-centroids}
\end{equation}

The phase-restricted face apertures are
\begin{equation}
A^{1\pm}_{i-\frac12,j}(t) := \langle \Sigma^{1\pm}_{i-\frac12,j}(t),1\rangle,
\qquad
A^{2\pm}_{i,j-\frac12}(t) := \langle \Sigma^{2\pm}_{i,j-\frac12}(t),1\rangle .
\label{eq:AxAy}
\end{equation}
When needed by the discrete operators, we also use the ``second kind'' spatial
moments (defined from the centroids above)
\begin{equation}
B^{1\pm}_{i,j}(t) := \left\langle \Sigma^{1\pm}_{j}\!\left(t,x^\pm_{i,j}(t)\right),1\right\rangle,
\qquad
B^{2\pm}_{i,j}(t) := \left\langle \Sigma^{2\pm}_{i}\!\left(t,y^\pm_{i,j}(t)\right),1\right\rangle,
\label{eq:BxBy}
\end{equation}
and the face-centered phase volumes
\begin{equation}
W^{1\pm}_{i-\frac12,j}(t)
:=
\left\langle
\Omega^{1\pm}_j\!\left(t,\left]x^\pm_{i-1,j}(t),\,x^\pm_{i,j}(t)\right[\right),
1\right\rangle,
\qquad
W^{2\pm}_{i,j-\frac12}(t)
:=
\left\langle
\Omega^{2\pm}_i\!\left(t,\left]y^\pm_{i,j-1}(t),\,y^\pm_{i,j}(t)\right[\right),
1\right\rangle.
\label{eq:WxWy}
\end{equation}

The moving-geometry extension requires that the same geometric information be
integrated in time over each time slab. This is the direct consequence
of applying Reynolds' transport theorem to control volumes that evolve through
$\Gamma(t)$ : conservation laws over
$[t_{n},t_{n+1}]$ involve space-time fluxes, which naturally introduce
time-integrated measures. Accordingly, we denote space-time (time-integrated)
moments by calligraphic letters. For any time-dependent quantity $M(t)$ defined
from spatial moments, we set
\begin{equation}
\mathcal{M}_{n+\frac12} := \int_{t_{n}}^{t_{n+1}} M(t)\,dt .
\label{eq:time_integral_generic}
\end{equation}

In particular, the time-integrated phase volumes and face apertures are
\begin{align}
\mathcal{V}^{\pm}_{n+\frac12,i,j}
&:= \int_{t_{n}}^{t_{n+1}} V^{\pm}_{i,j}(t)\,dt
 = \left\langle [t_n,t_{n+1}],\, V^\pm_{i,j}(t)\right\rangle,
\label{eq:time_integrated_VA1}
\\
\mathcal{A}^{1\pm}_{n+\frac12,i-\frac12,j}
&:= \int_{t_{n}}^{t_{n+1}} A^{1\pm}_{i-\frac12,j}(t)\,dt
 = \left\langle [t_n,t_{n+1}],\, A^{1\pm}_{i-\frac12,j}(t)\right\rangle,
\\
\mathcal{A}^{2\pm}_{n+\frac12,i,j-\frac12}
&:= \int_{t_{n}}^{t_{n+1}} A^{2\pm}_{i,j-\frac12}(t)\,dt
 = \left\langle [t_n,t_{n+1}],\, A^{2\pm}_{i,j-\frac12}(t)\right\rangle.
\label{eq:time_integrated_VA2}
\end{align}

with analogous definitions for the other Cartesian faces. These extensive
quantities satisfy $[\mathcal V]=\mathrm{m}^d\,\mathrm{s}$ and
$[\mathcal A]=\mathrm{m}^{d-1}\,\mathrm{s}$.

\begin{figure}[htbp]
  \centering
  \includegraphics[width=0.55\linewidth]{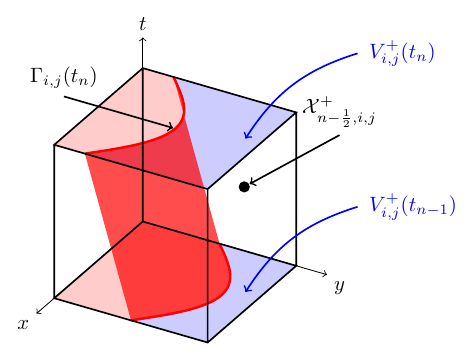}
  \caption{Space-time control volume over $[t_{n},t_{n+1}]$ illustrating the
  phase-restricted sub-volumes $V^+_{i,j}(t_{n})$ and $V^+_{i,j}(t_{n+1})$ and the
  swept interface within the time slab.}
  \label{fig:st-cell-volumes-only}
\end{figure}

Time-integrated centroids are defined from normalized first moments. In $d=2$,
\begin{align}
\mathcal{X}^{\pm}_{n+\frac12,i,j}
&:=
\begin{cases}
\displaystyle
\frac{\int_{t_{n}}^{t_{n+1}}\left\langle \Omega^\pm_{i,j}(t), x\right\rangle\,dt}
     {\mathcal{V}^{\pm}_{n+\frac12,i,j}}, &
\text{if }\mathcal{V}^{\pm}_{n+\frac12,i,j}\neq 0,\\[2.0ex]
x_i, & \text{otherwise},
\end{cases}
\label{eq:time_centroid_x}
\\
\mathcal{Y}^{\pm}_{n+\frac12,i,j}
&:=
\begin{cases}
\displaystyle
\frac{\int_{t_{n}}^{t_{n+1}}\left\langle \Omega^\pm_{i,j}(t), y\right\rangle\,dt}
     {\mathcal{V}^{\pm}_{n+\frac12,i,j}}, &
\text{if }\mathcal{V}^{\pm}_{n+\frac12,i,j}\neq 0,\\[2.0ex]
y_j, & \text{otherwise}.
\end{cases}
\label{eq:time_centroid_y}
\end{align}
These remain intensive, i.e.\ $[\mathcal X]=[\mathcal Y]=\mathrm{m}$.

The discrete operators further require time-integrated auxiliary measures (the
``moments of the second kind'') that are the
direct time-integrated counterparts of the static quantities introduced in the
companion paper. Using the space-time centroids above, we define
\begin{align}
\mathcal{B}^{1\pm}_{n+\frac12,i,j}
&:=
\int_{t_{n}}^{t_{n+1}}
\left\langle \Sigma^{1\pm}_{j}\!\left(t,\mathcal{X}^{\pm}_{n+\frac12,i,j}\right), 1\right\rangle\,dt,
\qquad
\mathcal{B}^{2\pm}_{n+\frac12,i,j}
:=
\int_{t_{n}}^{t_{n+1}}
\left\langle \Sigma^{2\pm}_{i}\!\left(t,\mathcal{Y}^{\pm}_{n+\frac12,i,j}\right), 1\right\rangle\,dt,
\label{eq:time_integrated_B}
\\
\mathcal{W}^{1\pm}_{n+\frac12,i-\frac12,j}
&:=
\int_{t_{n}}^{t_{n+1}}
\left\langle
\Bigl(\bigl(\mathcal{X}^{\pm}_{n+\frac12,i-1,j},\mathcal{X}^{\pm}_{n+\frac12,i,j}\bigr)
\times (y_{j-\frac12},y_{j+\frac12})\Bigr)\cap\Omega^\pm(t),\, 1
\right\rangle\,dt,
\label{eq:time_integrated_W1}
\\
\mathcal{W}^{2\pm}_{n+\frac12,i,j-\frac12}
&:=
\int_{t_{n}}^{t_{n+1}}
\left\langle
\Bigl((x_{i-\frac12},x_{i+\frac12})
\times \bigl(\mathcal{Y}^{\pm}_{n+\frac12,i,j-1},\mathcal{Y}^{\pm}_{n+\frac12,i,j}\bigr)\Bigr)\cap\Omega^\pm(t),\, 1
\right\rangle\,dt.
\label{eq:time_integrated_W2}
\end{align}

\begin{figure}[htbp]
  \centering
  \includegraphics[width=0.55\linewidth]{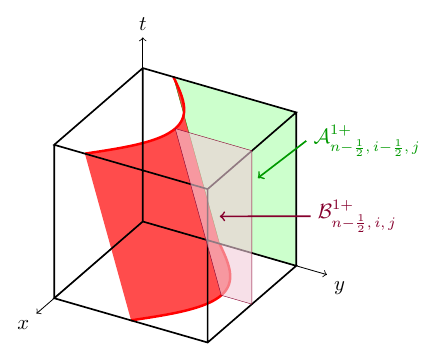}
  \caption{Space-time cell highlighting a time-integrated face aperture
  $\mathcal A^{1+}_{n+\frac12,i-\frac12,j}$ and an auxiliary measure of the
  second kind $\mathcal B^{1+}_{n+\frac12,i,j}$ used by the discrete operators.}
  \label{fig:st-face-ximinus}
\end{figure}

These definitions mirror the static construction; the only change is the
integration in time over the evolving phase domains. The resulting set
$\{\mathcal V,\mathcal A,\mathcal X,\mathcal Y,\mathcal B,\mathcal W\}$ provides
all geometric moments required by the space-time discrete operators and
remains well-defined even in the presence of fresh/dead cut cells within the
time slab.

Finally, we note that the calligraphic notation does not necessarily preserve
the physical dimensions of the corresponding spatial quantities: extensive
measures gain a factor of time (e.g.\ $[\mathcal V]=[V]\cdot\mathrm{s}$,
$[\mathcal A]=[A]\cdot\mathrm{s}$), whereas intensive measures such as centroids
retain their original dimension.

A table in Appendix~\ref{app:tablest} lists, for two spatial dimensions, the primary and secondary moments used in the discrete formulation.

\section{Space-time two-phase cut-cell method for moving domains}
\label{sec:moving-cutcell}

We now extend the static cut-cell operators to prescribed moving geometries.
To handle moving boundaries, we employ a space-time cut-cell formulation in
which time is treated as an additional coordinate: a $d$-dimensional Cartesian
mesh is viewed as a $(d+1)$-dimensional space-time mesh over each slab
$[t_{n},t_{n+1}]$ (e.g.\ 1D$\to$2D, 2D$\to$3D and 3D$\to$4D). In the remainder, we present the
two-dimensional case ($d=2$) for clarity but the construction extends directly
to $d=3$. This embedding turns each spatial cell into a space-time cut-cell,
defined as the $(d+1)$-dimensional polytope swept by the interface over one time
step. Its boundary includes the spatial faces at $t_{n}$ and $t_{n+1}$, the swept
lateral faces and the swept interface. 

Because the interface $\Gamma(t)$ evolves through the fixed Cartesian mesh, the
phase occupancy of a given cell may change during a time slab. For robust
assembly, we classify each spatial cell $\Omega_{i,j}$ by the behaviour of its
phase volumes at the slab endpoints and within the slab.

\paragraph{Regular cells.}
A cell is called \emph{regular} in phase $\pm$ over $[t_{n},t_{n+1}]$ if it remains pure
in that phase for all times in the slab, i.e.
$\Omega^\pm_{i,j}(t)=\Omega_{i,j}$ and $\Gamma_{i,j}(t)=\emptyset$ for all
$t\in[t_{n},t_{n+1}]$. In that case, the space-time control volume is the Cartesian
prism $\Omega_{i,j}\times[t_{n},t_{n+1}]$ and the geometric measures are obvious.

\paragraph{Cut cells and topology changes.}
A cell is said \emph{cut} over $[t_{n},t_{n+1}]$ if $\Gamma_{i,j}(t)\neq\emptyset$ for at
least one $t\in[t_{n},t_{n+1}]$. The corresponding phase volumes and face apertures
vary in time and must be handled through the time-integrated moments
$\mathcal{V}^\pm$, $\mathcal{A}^{\alpha\pm}$, etc. We further distinguish:

\begin{itemize}
\item \textbf{Persistent cut cells:} $\Gamma_{i,j}(t_{n})\neq\emptyset$ and
      $\Gamma_{i,j}(t_{n+1})\neq\emptyset$ (the cell is cut at both slab endpoints).

\item \textbf{Fresh cells for phase $\pm$:} $V^\pm_{i,j}(t_{n})=0$ and
      $V^\pm_{i,j}(t_{n+1})>0$. The phase appears in the cell during the slab.

\item \textbf{Dead cells for phase $\pm$:} $V^\pm_{i,j}(t_{n})>0$ and
      $V^\pm_{i,j}(t_{n+1})=0$. The phase disappears from the cell during the slab.
\end{itemize}

In all three cases, the space-time formulation provides a unique conservative
update because the exchange of extensive quantities is accounted for on the
complete space-time boundary (swept faces and swept interface) rather than by
separate ad hoc treatments at $t_{n}$ and $t_{n+1}$.
 In what follows,
we construct a space-time integrated divergence operator, accounting for
flux exchanges through the space-time boundary of the moving control volume,
and a compatible space-time gradient operator used to approximate diffusive
fluxes while preserving discrete consistency and the geometric conservation law.

\begin{figure}[!h]
\centering
\begin{subfigure}[t]{0.32\textwidth}
  \centering
  \includegraphics[width=\linewidth]{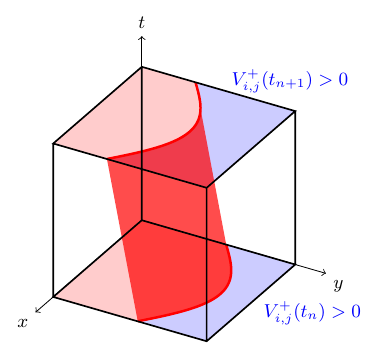}
  \caption{Persistent cut cell.}
  \label{fig:st-persistent}
\end{subfigure}\hfill
\begin{subfigure}[t]{0.32\textwidth}
  \centering
  \includegraphics[width=\linewidth]{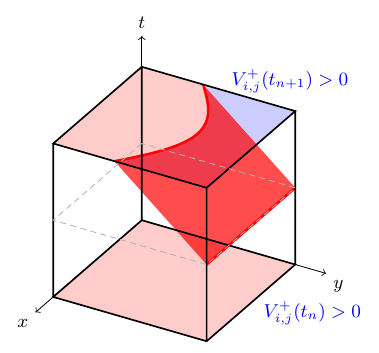}
  \caption{Fresh cell for phase $+$.}
  \label{fig:st-fresh}
\end{subfigure}\hfill
\begin{subfigure}[t]{0.32\textwidth}
  \centering
  \includegraphics[width=\linewidth]{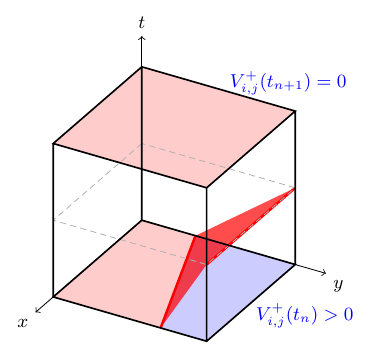}
  \caption{Dead cell for phase $+$.}
  \label{fig:st-dead}
\end{subfigure}

\caption{Space-time cut-cell taxonomy over one slab $[t_n,t_{n+1}]$: persistent cut cell (interface intersects the cell at both slab endpoints), fresh cell for phase $+$ (the $+$ phase appears during the slab) and dead cell for phase $+$ (the $+$ phase disappears during the slab).}
\label{fig:st-taxonomy}
\end{figure}

For each active phase cell, we define the space-time averaged bulk quantity
\begin{equation}
\widetilde{\Phi}^{\omega\pm}_{n+\sfrac{1}{2},i,j}
:=
\frac{\left\langle [t_{n},t_{n+1}],\,\langle \Omega_{i,j}^{\pm}(t),\phi^{\pm}(t)\rangle \right\rangle}
{\mathcal{V}_{n+\sfrac{1}{2},i,j}^{\pm}} \quad \mathrm{where} \quad \mathcal V ^ \pm _ {n+\sfrac12,i, j} \ne 0,
\label{eq:st_bulk_avg}
\end{equation}
The discrete (time-slab endpoint) bulk unknowns remain those of the static
formulation,
\begin{equation}
\Phi^{\omega\pm}_{n,i,j} := \Phi^{\omega\pm}_{i,j}(t_{n}).
\label{eq:bulk_endpoint}
\end{equation}
where we recall the definition of the spatial bulk average
\begin{equation*}
    \Phi _ {i, j} ^ {\omega \pm} \left ( t \right ) := 
     \left \langle \Omega ^ \pm _ {i, j}, \phi^\pm \left ( t \right ) \right \rangle / V ^ \pm _ {i, j} \quad \mathrm{where} \quad V ^ \pm _ {i, j} \ne 0,
\end{equation*}
On interfacial (mixed) cells, we similarly introduce space-time averaged
interfacial traces, defined separately on each side of $\Gamma(t)$,
\begin{equation}
\Phi^{\gamma\pm}_{n+\frac12,i,j}
:=
\frac{
\displaystyle
\int_{t_{n}}^{t_{n+1}}\!\!\left\langle \Gamma_{i,j}(t),\,\phi^{\pm}(t)\right\rangle\,dt
}{
\displaystyle
\int_{t_{n}}^{t_{n+1}}\!\!\int_{\Gamma_{i,j}(t)} 1\,dS\,dt
}.
\label{eq:st_interface_avg}
\end{equation}
These interfacial quantities enter only through the same two-fluid coupling
laws as in the static case.

To ensure that these discrete events are detected and treated within a single
time step, we enforce an interface CFL-like constraint: the interface must not
cross more than one cell in the normal direction during one update. This restriction prevents ``skipped'' fresh/dead
transitions and guarantees that the time-integrated moments
$\mathcal{V}^\pm_{n+\frac12,i,j}$ and $\mathcal{A}^{\alpha\pm}_{n+\frac12,\cdot}$
capture all topological changes occurring in the slab.

\subsection{Semi-discrete and discrete bulk equation}
\label{sec:sd-bulk-moving}

The starting point is Reynolds' transport theorem applied to the time-evolving
phase cell $\Omega^\pm_{i,j}(t)$,
\begin{equation}
\frac{d}{dt}\left\langle \Omega^\pm_{i,j}(t),\, C^\pm \phi^\pm \right\rangle
=
\left\langle \Omega^\pm_{i,j}(t),\, \partial_t(C^\pm\phi^\pm)\right\rangle
+
\int_{\partial\Omega^\pm_{i,j}(t)} C^\pm\phi^\pm\,u^\pm\,dS,
\label{eq:rtt-moving}
\end{equation}
where $u^\pm$ denotes the normal velocity of the moving boundary element.
By definition of $\Omega^\pm_{i,j}(t)$, the only
moving part of $\partial\Omega^\pm_{i,j}(t)$ is the interface segment
$\Gamma_{i,j}(t)$. Cartesian cell faces are fixed. With the choice
$\mathbf n=\mathbf n^-$ (red phase reference), this yields
\[
u^\pm=
\begin{cases}
\mp\,w, & \text{on }\Gamma_{i,j}(t),\\
0, & \text{otherwise},
\end{cases}
\]

Using the bulk balance equation \eqref{eq:bulk-balance}, the divergence theorem, 
we obtain the exact semi-discrete balance
\begin{equation}
C^\pm \frac{d}{dt}\!\left[\,V^\pm_{i,j}(t)\,\Phi^{\omega\pm}_{i,j}(t)\right]
+
\int_{\partial\Omega^\pm_{i,j}(t)} \mathbf q^\pm(t)\cdot d\mathbf S
\ \mp\
\int_{\Gamma_{i,j}(t)} C^\pm\phi^\pm(t)\,w\,dS
=
\left\langle \Omega^\pm_{i,j}(t),\, r^\pm(t)\right\rangle,
\label{eq:sd-raw}
\end{equation}
For convenience, we introduce the phase source average
\[
R^\pm_{i,j}(t)
:=
\frac{\left\langle \Omega^\pm_{i,j}(t),\, r^\pm(t)\right\rangle}{V^\pm_{i,j}(t)}
\qquad\text{when }V^\pm_{i,j}(t)\neq 0.
\]

Integrating \eqref{eq:sd-raw} over the time slab $[t_{n},t_{n+1}]$ gives the
(space-time) fully discrete bulk equation:
\begin{multline}
C^\pm\Bigl(
V^\pm_{n+1,i,j}\,\Phi^{\omega\pm}_{n+1,i,j}
-
V^\pm_{n,i,j}\,\Phi^{\omega\pm}_{n,i,j}
\Bigr)
+
\int_{t_{n}}^{t_{n+1}}\!\!\int_{\partial\Omega^\pm_{i,j}(t)}
\mathbf q^\pm(t)\cdot d\mathbf S\,dt
\\
\mp
\int_{t_{n}}^{t_{n+1}}\!\!\int_{\Gamma_{i,j}(t)} C^\pm\phi^\pm(t)\,w\,dS\,dt
=
\mathcal V^\pm_{n+\frac12,i,j}\,\mathcal R^\pm_{n+\frac12,i,j},
\label{eq:st_bulk}
\end{multline}
with the
space-time source average
\[
\mathcal R^\pm_{n+\frac12,i,j}
:=
\frac{
\int_{t_{n}}^{t_{n+1}}\left\langle \Omega^\pm_{i,j}(t),\, r^\pm(t)\right\rangle\,dt
}{
\mathcal V^\pm_{n+\frac12,i,j}
}
\qquad \text{when }\mathcal V^\pm_{n+\frac12,i,j}\neq 0.
\]

\subsubsection{Space-time-integrated divergence operator}
\label{sec:st-div}

As in the static case, the time-integrated boundary flux in
\eqref{eq:st_bulk} is split into contributions through Cartesian faces (exchange
with neighboring control volumes in the same phase) and through the interface:
\begin{equation}
\int_{t_{n}}^{t_{n+1}}\!\!\int_{\partial\Omega^\pm_{i,j}(t)}
\mathbf q^\pm\cdot d\mathbf S\,dt
=
\int_{t_{n}}^{t_{n+1}}\!\!\int_{\partial\Omega^\pm_{i,j}(t)\setminus\Gamma_{i,j}(t)}
\mathbf q^\pm\cdot d\mathbf S\,dt
+
\int_{t_{n}}^{t_{n+1}}\!\!\int_{\Gamma_{i,j}(t)}
\mathbf q^\pm\cdot d\mathbf S\,dt.
\label{eq:gauss-st}
\end{equation}

Both contributions are discretized using the same volume-integrated divergence
operators as in the static formulation, the only change
being that the geometric inputs are now time-integrated:
\[
\sum_{\alpha=1}^d
\mathcal V^\pm_{n+\frac12,i,j}\,
\operatorname{div}^{\alpha\omega}_{i,j}\!\left(
\mathcal A^{\alpha\pm}_{n+\frac12},\,\mathcal Q^{\alpha\pm}_{n+\frac12}
\right)
\qquad\text{and}\qquad
\sum_{\alpha=1}^d
\mathcal V^\pm_{n+\frac12,i,j}\,
\operatorname{div}^{\alpha\gamma}_{i,j}\!\left(
\mathcal A^{\alpha\pm}_{n+\frac12},\,\mathcal B^{\alpha\pm}_{n+\frac12},\,\mathcal Q^{\alpha\pm}_{n+\frac12}
\right),
\]
for the Cartesian-face and interface parts, respectively.

Accordingly, we define the time-integrated diffusive fluxes by space-time
averaging over the corresponding time-dependent phase-restricted faces. In 2D,
the $x$- and $y$-fluxes are
\begin{align}
\mathcal Q^{1\pm}_{n+\frac12,i+\frac12,j}
&\simeq
\frac{
\displaystyle
\int_{t_{n}}^{t_{n+1}}\left\langle \Sigma^{1\pm}_{i+\frac12,j}(t),\, q^{1\pm}(t)\right\rangle\,dt
}{
\mathcal A^{1\pm}_{n+\frac12,i+\frac12,j}
},
\qquad
\text{when }\mathcal A^{1\pm}_{n+\frac12,i+\frac12,j}\neq 0,
\label{eq:flux-tx}
\\
\mathcal Q^{2\pm}_{n+\frac12,i,j+\frac12}
&\simeq
\frac{
\displaystyle
\int_{t_{n}}^{t_{n+1}}\left\langle \Sigma^{2\pm}_{i,j+\frac12}(t),\, q^{2\pm}(t)\right\rangle\,dt
}{
\mathcal A^{2\pm}_{n+\frac12,i,j+\frac12}
},
\qquad
\text{when }\mathcal A^{2\pm}_{n+\frac12,i,j+\frac12}\neq 0,
\label{eq:flux-ty}
\end{align}

\begin{figure}[!h]
\centering
\includegraphics[width=0.68\textwidth]{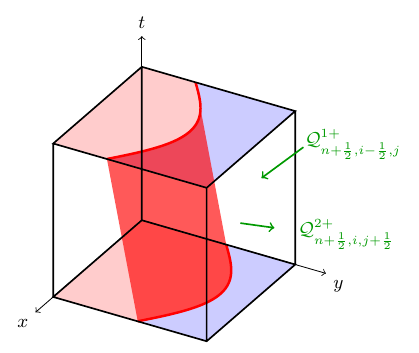}
\caption{Space-time cut-cell representation of the time-integrated diffusive fluxes through Cartesian faces over one slab $[t_n,t_{n+1}]$. Only two fluxes are shown for clarity: the $x$-face flux $\mathcal Q^{1+}_{n+\frac12,\,i-\frac12,\,j}$ and the $y$-face flux $\mathcal Q^{2+}_{n+\frac12,\,i,\,j+\frac12}$, defined as space-time averages over the corresponding phase-restricted swept faces.}
\label{fig:st-fluxes}
\end{figure}

\subsubsection{Geometric conservation law}
\label{sec:gcl}

A fundamental requirement of moving-boundary finite-volume schemes
is the exact preservation of constant states. Consider the particular case where
$C^\pm$ and $\phi^\pm$ are constants, so that $\mathbf q^\pm\equiv \mathbf 0$ and
$r^\pm\equiv 0$. The bulk balance \eqref{eq:bulk-balance} then reduces to the
identity $0=0$ but integrating it over the evolving phase cell
$\Omega^\pm_{i,j}(t)$ over one time slab yields the purely geometric relation
\begin{equation}
V^\pm_{n+1,i,j}-V^\pm_{n,i,j}
\ \mp\
\int_{t_{n}}^{t_{n+1}}\!\!\int_{\Gamma_{i,j}(t)} w\,dS\,dt
=0,
\label{eq:gcl}
\end{equation}
which is the geometric conservation law (GCL) in the present cut-cell setting.
A discrete scheme that satisfies \eqref{eq:gcl} reproduces constant solutions
exactly and prevents spurious source terms induced solely by geometric motion ~\cite{farhat_discrete_2001}.

To approximate the interfacial sweeping term in \eqref{eq:st_bulk}, we apply the
space-time quadrature consistent with the definition of the interfacial trace
\eqref{eq:st_interface_avg}:
\[
\int_{t_{n}}^{t_{n+1}}\!\!\int_{\Gamma_{i,j}(t)} C^\pm\phi^\pm(t)\,w\,dS\,dt
\;\simeq\;
C^\pm\,\Phi^{\gamma\pm}_{n+\frac12,i,j}\,
\int_{t_{n}}^{t_{n+1}}\!\!\int_{\Gamma_{i,j}(t)} w\,dS\,dt.
\]
Using the GCL \eqref{eq:gcl}, this becomes
\begin{equation}
\int_{t_{n}}^{t_{n+1}}\!\!\int_{\Gamma_{i,j}(t)} C^\pm\phi^\pm(t)\,w\,dS\,dt
\;\simeq\;
\mp\,C^\pm\,\Phi^{\gamma\pm}_{n+\frac12,i,j}\,
\bigl(V^\pm_{n+1,i,j}-V^\pm_{n,i,j}\bigr).
\label{eq:sweep_approx_gcl}
\end{equation}

Substituting \eqref{eq:sweep_approx_gcl} together with the space-time-integrated
divergence discretization of Section~\ref{sec:st-div} into \eqref{eq:st_bulk} yields
the discrete bulk balance in conservative space-time form:
\begin{multline}
C^\pm\Bigl(
V^\pm_{n+1,i,j}\,\Phi^{\omega\pm}_{n+1,i,j}
-
V^\pm_{n,i,j}\,\Phi^{\omega\pm}_{n,i,j}
\Bigr)
\\
+
\sum_{\alpha=1}^d
\Bigl[
\mathcal V^\pm_{n+\frac12,i,j}\,
\operatorname{div}^{\alpha\omega}_{i,j}\!\left(
\mathcal A^{\alpha\pm}_{n+\frac12},\,\mathcal Q^{\alpha\pm}_{n+\frac12}
\right)
+
\mathcal V^\pm_{n+\frac12,i,j}\,
\operatorname{div}^{\alpha\gamma}_{i,j}\!\left(
\mathcal A^{\alpha\pm}_{n+\frac12},\,\mathcal B^{\alpha\pm}_{n+\frac12},\,\mathcal Q^{\alpha\pm}_{n+\frac12}
\right)
\Bigr]
\\
+ \,C^\pm\,\Phi^{\gamma\pm}_{n+\frac12,i,j}\,
\bigl(V^\pm_{n+1,i,j}-V^\pm_{n,i,j}\bigr)
=
\mathcal V^\pm_{n+\frac12,i,j}\,\mathcal R^\pm_{n+\frac12,i,j}.
\label{eq:st-bulk-divergence}
\end{multline}
Equation \eqref{eq:st-bulk-divergence} reduces to the static cut-cell balance
when the geometry is fixed (all time-integrated measures reduce to
$\mathcal A\!=\!\Delta t_n\,A$, $\mathcal V\!=\!\Delta t_n\,V$ and the volume
difference vanishes).

\subsubsection{Space-time-integrated gradient operator}
\label{sec:st-grad}

The construction of the diffusive fluxes in \eqref{eq:st-bulk-divergence}
follows the same principles as in the static formulation but with space-time
geometric weights. In particular, the constitutive law
$\mathbf q^\pm=-K^\pm\nabla\phi^\pm$ requires a discrete gradient operator that
is compatible with the space-time divergence operators.

We therefore define the space-time gradient operator by reusing the static
bulk and interfacial gradient components, replacing $(A,B,W)$ by their
time-integrated counterparts $(\mathcal A,\mathcal B,\mathcal W)$ and using the
space-time averaged unknowns $\widetilde\Phi^{\omega\pm}$ and
$\Phi^{\gamma\pm}_{n+\frac12}$:
\[
\operatorname{grad}^{1\omega}_{i-\frac12,j}
\!\left(\mathcal B^{1\pm}_{n+\frac12},\,\mathcal W^{1\pm}_{n+\frac12},\,\widetilde\Phi^{\omega\pm}_{n+\frac12}\right),
\qquad
\operatorname{grad}^{1\gamma}_{i-\frac12,j}
\!\left(\mathcal A^{1\pm}_{n+\frac12},\,\mathcal B^{1\pm}_{n+\frac12},\,\mathcal W^{1\pm}_{n+\frac12},\,\Phi^{\gamma\pm}_{n+\frac12}\right),
\]
with analogous expressions in the other coordinate directions. The space-time
diffusive flux through an $x$-oriented face then reads
\begin{equation}
\mathcal Q^{1\pm}_{n+\frac12,i-\frac12,j}
=
-\,K^\pm\,
\Big[
\operatorname{grad}^{1\omega}_{i-\frac12,j}
\!\left(\mathcal B^{1\pm}_{n+\frac12},\,\mathcal W^{1\pm}_{n+\frac12},\,\widetilde\Phi^{\omega\pm}_{n+\frac12}\right)
+
\operatorname{grad}^{1\gamma}_{i-\frac12,j}
\!\left(\mathcal A^{1\pm}_{n+\frac12},\,\mathcal B^{1\pm}_{n+\frac12},\,\mathcal W^{1\pm}_{n+\frac12},\,\Phi^{\gamma\pm}_{n+\frac12}\right)
\Big],
\label{eq:st-constitutive}
\end{equation}
and similarly for $\mathcal Q^{2\pm}_{n+\frac12,i,j-\frac12}$ in the $y$
direction. This construction reduces exactly to
the static one when the geometry is fixed and provides a space-time consistent
approximation of the diffusive term in \eqref{eq:st-bulk-divergence}.

\subsubsection{Fresh and dead cells: topology change}
\label{sec:freshdead}

The fully discrete space-time balances
\eqref{eq:st-bulk-divergence}-\eqref{eq:st-constitutive} apply directly to
regular cells and persistent cut cells, i.e.\ configurations for which both
endpoint phase volumes are nonzero,
$V^\pm_{n,i,j}>0$ and $V^\pm_{n+1,i,j}>0$.
In that case, the space-time bulk state used by the constitutive law is closed
by the standard $\theta$-scheme between $t_n$ and $t_{n+1}$.

When the interface motion creates or removes a phase within a cell during the
slab, i.e.\ for fresh or dead cut cells, the space-time flux
evaluation still requires a well-defined slab state
$\widetilde{\Phi}^{\omega\pm}_{n+\frac12,i,j}$ even though one of the endpoint
unknowns is absent. We therefore adopt the following state-dependent closure:
\begin{equation}
\widetilde{\Phi}^{\omega\pm}_{n+\frac12,i,j} =
\begin{cases}
\Phi^{\omega\pm}_{n+1,i,j},
& V^\pm_{n,i,j}=0,\; V^\pm_{n+1,i,j}>0
\quad \text{(fresh cell)},\\[4pt]
{\Phi}^{\omega\pm}_{\mathrm{loc},i,j},
& V^\pm_{n,i,j}>0,\; V^\pm_{n+1,i,j}=0
\quad \text{(dead cell)},\\[4pt]
(1-\theta)\,\Phi^{\omega\pm}_{n,i,j}+\theta\,\Phi^{\omega\pm}_{n+1,i,j},
& V^\pm_{n,i,j}>0,\; V^\pm_{n+1,i,j}>0
\quad \text{(persistent/regular cell)}.
\end{cases}
\label{eq:tildephi-definition}
\end{equation}
In persistent/regular cells, \eqref{eq:tildephi-definition} reduces to the usual
time interpolation and is passed to the space-time gradient operator in the
constitutive relation \eqref{eq:st-constitutive}. In transitioning cells, it
provides a consistent slab state so that the face fluxes remain well defined
and global conservation over $[t_n,t_{n+1}]$ is preserved.

\paragraph{Dead cells.}
For dead cells ($V^\pm_{n+1,i,j}=0$), no bulk unknown is advanced at $t_{n+1}$
because the phase becomes inactive in $\Omega_{i,j}$.
The closure value ${\Phi}^{\omega\pm}_{\mathrm{loc},i,j}$ is therefore computed
locally by enforcing that the space-time balance
\eqref{eq:st-bulk-divergence} holds on the disappearing control volume, with
all face fluxes assembled using $\widetilde\Phi^{\omega\pm}$ and the space-time
geometry:
\begin{multline}
-\,C^\pm\,V^\pm_{n,i,j}\,\Phi^{\omega\pm}_{n,i,j}
+
\sum_{\alpha=1}^{d}\Big[
\mathcal V^\pm_{n+\frac12,i,j}\,
\operatorname{div}^{\alpha\omega}_{i,j}\!\Big(
\mathcal A^{\alpha\pm}_{n+\frac12},\,
\mathcal Q^{\alpha\pm}_{n+\frac12}(\widetilde{\Phi}^{\omega\pm})
\Big)
\\
+
\mathcal V^\pm_{n+\frac12,i,j}\,
\operatorname{div}^{\alpha\gamma}_{i,j}\!\Big(
\mathcal A^{\alpha\pm}_{n+\frac12},\,
\mathcal B^{\alpha\pm}_{n+\frac12},\,
\mathcal Q^{\alpha\pm}_{n+\frac12}(\widetilde{\Phi}^{\omega\pm})
\Big)
\Big]
\\
-\,C^\pm\,\Phi^{\gamma\pm}_{n+\frac12,i,j}\,V^\pm_{n,i,j}
=
\mathcal V^\pm_{n+\frac12,i,j}\,\mathcal R^\pm_{n+\frac12,i,j}.
\label{eq:dead-closure}
\end{multline}
Since $V^\pm_{n+1,i,j}=0$, \eqref{eq:dead-closure} yields a single scalar
linear constraint for ${\Phi}^{\omega\pm}_{\mathrm{loc},i,j}$ because the fluxes
$\mathcal Q^{\alpha\pm}_{n+\frac12}$ depend linearly on the slab states through
\eqref{eq:st-constitutive}. In practice, \eqref{eq:dead-closure} is assembled as
\[
a^\pm_{i,j}\,{\Phi}^{\omega\pm}_{\mathrm{loc},i,j} = b^\pm_{i,j},
\]
where $a^\pm_{i,j}$ and $b^\pm_{i,j}$ gather, respectively, the coefficients and
known terms produced by the (linear) divergence-gradient-constitutive
composition when only the local dead-cell state is left unknown. The resulting
${\Phi}^{\omega\pm}_{\mathrm{loc},i,j}$ is used only to evaluate fluxes in
the current slab and is not carried to the next time level.

\paragraph{Fresh cells.}
For newly created cells ($V^\pm_{n,i,j}=0$, $V^\pm_{n+1,i,j}>0$), we set
\eqref{eq:tildephi-definition} with $\widetilde{\Phi}^{\omega\pm}_{n+\frac12,i,j}
=\Phi^{\omega\pm}_{n+1,i,j}$. The slab balance then specializes to
\begin{multline}
C^\pm\,V^\pm_{n+1,i,j}\,\Phi^{\omega\pm}_{n+1,i,j}
+
\sum_{\alpha=1}^{d}\Big[
\mathcal V^\pm_{n+\frac12,i,j}\,
\operatorname{div}^{\alpha\omega}_{i,j}\!\Big(
\mathcal A^{\alpha\pm}_{n+\frac12},\,
\mathcal Q^{\alpha\pm}_{n+\frac12}(\widetilde{\Phi}^{\omega\pm})
\Big)
\\
+
\mathcal V^\pm_{n+\frac12,i,j}\,
\operatorname{div}^{\alpha\gamma}_{i,j}\!\Big(
\mathcal A^{\alpha\pm}_{n+\frac12},\,
\mathcal B^{\alpha\pm}_{n+\frac12},\,
\mathcal Q^{\alpha\pm}_{n+\frac12}(\widetilde{\Phi}^{\omega\pm})
\Big)
\Big]
\\
-\,C^\pm\,\Phi^{\gamma\pm}_{n+\frac12,i,j}\,V^\pm_{n+1,i,j}
=
\mathcal V^\pm_{n+\frac12,i,j}\,\mathcal R^\pm_{n+\frac12,i,j},
\label{eq:fresh-balance}
\end{multline}
so no additional local unknown is introduced: fluxes entering the fresh cell are
computed using the initialized slab state, which activates the new control
volume consistently.

Summing the discrete balances over all cells (regular, persistent, fresh and
dead) and adding the discrete interface balance of Section~\ref{sec:st-interface-balance}
cancels all internal face contributions, leaving only volumetric sources. Global
conservation is therefore maintained across topology changes.

\subsection{Space-time interface balance}
\label{sec:st-interface-balance}

The interface equations are discretized as in the static formulation but the
space-time setting introduces the additional sweeping contribution associated
with interface motion. On each interfacial segment $\Gamma_{i,j}(t)$, the
instantaneous surface balance \eqref{eq:surface-balance} reads
\begin{equation}
\left\llbracket \mathbf q\cdot\mathbf n - C\,\phi\,w \right\rrbracket = 0,
\label{eq:surface-balance-moving}
\end{equation}

Integrating \eqref{eq:surface-balance-moving} over the slab $[t_n,t_{n+1}]$ and
discretizing the resulting surface integrals using the space-time geometric
operators yields the space-time interface balance. Using the same
space-time-interfacial quadrature as in Section~\ref{sec:gcl}, the sweeping term
can be written in terms of the phase-volume change over the slab, leading to
\begin{multline}
\sum_{\alpha=1}^{d}\Big[
\mathcal V^{+}_{n+\frac12,i,j}\,
\operatorname{div}^{\alpha\gamma}_{i,j}\!\left(
\mathcal A^{\alpha +}_{n+\frac12},\,\mathcal B^{\alpha +}_{n+\frac12},\,\mathcal Q^{\alpha +}_{n+\frac12}
\right)
-
\mathcal V^{-}_{n+\frac12,i,j}\,
\operatorname{div}^{\alpha\gamma}_{i,j}\!\left(
\mathcal A^{\alpha -}_{n+\frac12},\,\mathcal B^{\alpha -}_{n+\frac12},\,\mathcal Q^{\alpha -}_{n+\frac12}
\right)
\Big]
\\
-
\left\llbracket C\,\Phi^{\gamma\pm}_{n+\frac12,i,j}\right\rrbracket\,
\left(
V^{-}_{n+1,i,j}-V^{-}_{n,i,j}
\right)
= 0,
\label{eq:st-surface-balance}
\end{multline}
where the last term expresses the space-time sweeping flux in terms of the
phase-volume variation (here written using the reference phase ``$-$'').
Equivalent expressions are obtained by using the ``$+$'' phase, since
$V^-_{n+1,i,j}-V^-_{n,i,j}=-(V^+_{n+1,i,j}-V^+_{n,i,j})$.

The interfacial closure (continuity / Henry law / prescribed jump) is unchanged
from the static case and is imposed at the same intermediate time level
$t_{n+\frac12}$:
\begin{equation}
\Phi^{\gamma +}_{n+\frac12,i,j}
-
\lambda\,\Phi^{\gamma -}_{n+\frac12,i,j}
=
\mathcal F_{n+\frac12,i,j},
\label{eq:st-continuity}
\end{equation}
where $\mathcal F$ denotes the space-time averaged interfacial source term.

In the particular case of planar, mesh-aligned interfaces with $w=0$ (no
sweeping), \eqref{eq:st-surface-balance}-\eqref{eq:st-continuity} reduce to the
static interface relations.

\subsection{Block structure of the linear system}
\label{sec:block_system_st}

As in the static formulation, the continuum model is linear and the fully
discrete space-time equations over a slab $[t_n,t_{n+1}]$ can be written in
sparse matrix-vector form. Even for $\theta=0$ (explicit Euler), the unknowns
at time level $n+1$ remain coupled through the interfacial balance and closure.
At each step, the numerical update therefore amounts to solving a sparse linear
system coupling bulk and interfacial unknowns.

We eliminate the space-time diffusive fluxes $\mathcal Q^{\alpha\pm}$ using the
constitutive closure \eqref{eq:st-constitutive}, leaving the bulk and interfacial
unknowns as the remaining degrees of freedom. We use the ordering
\[
\mathbf x_{n+1}
=
\bigl(
\Phi^{\omega-}_{n+1},\;
\Phi^{\omega+}_{n+1},\;
\Phi^{\gamma-}_{n+\frac12},\;
\Phi^{\gamma+}_{n+\frac12}
\bigr).
\]

We introduce diagonal mass-like matrices
\begin{equation}
M^\pm_{\mathrm{st}}
:=
\frac{C^\pm}{\Delta t_n}\,\operatorname{diag}\!\bigl(V^\pm_{n+1}\bigr),
\label{eq:Mst}
\end{equation}
In addition, the sweeping (interface-advection) contribution in
\eqref{eq:st-bulk-divergence} introduces a diagonal coupling between the bulk
equations and the interfacial unknowns. We collect it into
\begin{equation}
A^\pm_{\mathrm{adv}}
:=
\mp C^\pm\,\operatorname{diag}\!\bigl(V^\pm_{n+1}-V^\pm_{n}\bigr),
\label{eq:Aadv}
\end{equation}
which multiplies $\Phi^{\gamma\pm}_{n+\frac12}$ in the bulk balance of phase $\pm$.

Let $\operatorname{div}^{\alpha\omega}$ and $\operatorname{div}^{\alpha\gamma}$
denote the bulk and interfacial contributions to the divergence operator used in
\eqref{eq:st-bulk-divergence} and let
$\operatorname{grad}^{\alpha\omega}$ and $\operatorname{grad}^{\alpha\gamma}$
denote the corresponding gradient contributions used in \eqref{eq:st-constitutive}.
Eliminating $\mathcal Q^{\alpha\pm}$ yields space-time diffusion operators built
from the space-time moments $(\mathcal V,\mathcal A,\mathcal B,\mathcal W)$:
\begin{align}
L^{\omega\omega,\pm}_{\mathrm{st}}
&:=
-\sum_{\alpha}
\partial_{\mathcal Q}\!\Bigl[
\mathcal V^\pm\,
\operatorname{div}^{\alpha\omega}\bigl(\mathcal A^{\alpha\pm}_{n+\frac12},\cdot\bigr)
\Bigr]\;
K^\pm\;
\partial_{\Phi}\,
\operatorname{grad}^{\alpha\omega}\bigl(\mathcal B^{\alpha\pm}_{n+\frac12},
\mathcal W^{\alpha\pm}_{n+\frac12},\cdot\bigr),
\label{eq:Lww_st}\\
L^{\omega\gamma,\pm}_{\mathrm{st}}
&:=
-\sum_{\alpha}
\partial_{\mathcal Q}\!\Bigl[
\mathcal V^\pm\,
\operatorname{div}^{\alpha\omega}\bigl(\mathcal A^{\alpha\pm}_{n+\frac12},\cdot\bigr)
\Bigr]\;
K^\pm\;
\partial_{\Phi}\,
\operatorname{grad}^{\alpha\gamma}\bigl(\mathcal A^{\alpha\pm}_{n+\frac12},
\mathcal B^{\alpha\pm}_{n+\frac12},
\mathcal W^{\alpha\pm}_{n+\frac12},\cdot\bigr),
\label{eq:Lwg_st}\\
L^{\gamma\omega,\pm}_{\mathrm{st}}
&:=
-\sum_{\alpha}
\partial_{\mathcal Q}\!\Bigl[
\mathcal V^\pm\,
\operatorname{div}^{\alpha\gamma}\bigl(\mathcal A^{\alpha\pm}_{n+\frac12},
\mathcal B^{\alpha\pm}_{n+\frac12},\cdot\bigr)
\Bigr]\;
K^\pm\;
\partial_{\Phi}\,
\operatorname{grad}^{\alpha\omega}\bigl(\mathcal B^{\alpha\pm}_{n+\frac12},
\mathcal W^{\alpha\pm}_{n+\frac12},\cdot\bigr),
\label{eq:Lgw_st}\\
L^{\gamma\gamma,\pm}_{\mathrm{st}}
&:=
-\sum_{\alpha}
\partial_{\mathcal Q}\!\Bigl[
\mathcal V^\pm\,
\operatorname{div}^{\alpha\gamma}\bigl(\mathcal A^{\alpha\pm}_{n+\frac12},
\mathcal B^{\alpha\pm}_{n+\frac12},\cdot\bigr)
\Bigr]\;
K^\pm\;
\partial_{\Phi}\,
\operatorname{grad}^{\alpha\gamma}\bigl(\mathcal A^{\alpha\pm}_{n+\frac12},
\mathcal B^{\alpha\pm}_{n+\frac12},
\mathcal W^{\alpha\pm}_{n+\frac12},\cdot\bigr).
\label{eq:Lgg_st}
\end{align}
Here $\partial_{\mathcal Q}$ and $\partial_{\Phi}$ denote Jacobians of the
divergence and gradient operators with respect to fluxes and unknowns,
respectively. Since the discrete operators are linear in $\mathcal Q$ and $\Phi$,
these Jacobians reduce to constant sparse matrices.

Collecting the two bulk balances, the space-time interface balance and the
interfacial closure, we obtain the block system
\begin{multline}
\left[
\begin{array}{c|c|cc}
M^-_{\mathrm{st}} + \theta L^{\omega\omega,-}_{\mathrm{st}} & 0
  & \theta L^{\omega\gamma,-}_{\mathrm{st}} - A_{\mathrm{adv}}^- & 0 \\ \hline
0 & M^+_{\mathrm{st}} + \theta L^{\omega\omega,+}_{\mathrm{st}}
  & 0 & \theta L^{\omega\gamma,+}_{\mathrm{st}} - A_{\mathrm{adv}}^+ \\ \hline
-\theta L^{\gamma\omega,-}_{\mathrm{st}} &
 \theta L^{\gamma\omega,+}_{\mathrm{st}} &
 - L^{\gamma\gamma,-}_{\mathrm{st}} + A_{\mathrm{adv}}^- & L^{\gamma\gamma,+}_{\mathrm{st}} - A_{\mathrm{adv}}^+\\
0 & 0 & -\lambda I & I
\end{array}
\right]
\left[
\begin{array}{c}
\Phi^{\omega-}_{n+1} \\
\Phi^{\omega+}_{n+1} \\
\Phi^{\gamma-}_{n+\frac12} \\
\Phi^{\gamma+}_{n+\frac12}
\end{array}
\right]
=
\mathrm{RHS}_{\mathrm{st}},
\label{eq:st-matrix-system}
\end{multline}
where $\mathrm{RHS}_{\mathrm{st}}$ gathers all known contributions from the
previous time level (e.g.\ $\Phi^{\omega\pm}_{n}$), space-time source terms and
any prescribed interfacial forcing.

The horizontal and vertical separators highlight the same arrow-type coupling as
in the static case: two decoupled bulk blocks coupled only through the interface
unknowns.

In the limit of a fixed interface ($V^\pm_{n+1}=V^\pm_n$, hence
$A^\pm_{\mathrm{adv}}=0$ and $\mathcal A\to A\,\Delta t_n$,
$\mathcal B\to B\,\Delta t_n$, $\mathcal W\to W\,\Delta t_n$,
$\mathcal V\to V\,\Delta t_n$), \eqref{eq:st-matrix-system} reduces exactly to
the static block formulation.

\section{Numerical validation in moving domains}
\label{sec:nummotion}

This section assesses the accuracy and robustness of the proposed space-time
cut-cell method in the presence of prescribed moving geometries. All tests are
performed on a fixed Cartesian background grid while the physical domain
evolves through a sharp interface $\Gamma(t)$, continuously creating and
removing cut cells within each time slab. The numerical study targets three
complementary objectives: (i) verifying the correctness of the space-time
geometric moments and quadratures used by the operators, including the handling
of fresh/dead cells; (ii) validating the full diffusion solver on classical
moving-boundary benchmarks in two and three dimensions; and (iii) demonstrating
that the two-phase coupling remains accurate and conservative when the
interface sweeps across the mesh.

Let $\phi^{\rm ex}(\mathbf x,t)$ denote the exact solution. Errors are evaluated from
cell-averaged unknowns at selected times $t_n$ (typically the final time),
and reported separately over regular cells and cut cells.
The index sets are  defined by
\[
\mathcal I_{\rm reg}=\{(i,j):\ \Gamma_{i,j}=\emptyset\},\qquad
\mathcal I_{\rm cut}=\{(i,j):\ \Gamma_{i,j}\neq\emptyset\},\qquad
\mathcal I_{\rm all}=\mathcal I_{\rm reg}\cup \mathcal I_{\rm cut}.
\]

For any subset $S\in\{\mathrm{reg},\mathrm{cut},\mathrm{all}\}$, we define the
discrete $L^2$ error norm as
\begin{equation}
\|e(t_n)\|_{2,S}^{\rm}
=
\left(
\frac{\displaystyle
\sum_{(i,j)\in\mathcal I_S}
V_{i,j}\,
\bigl|\Phi^{\rm \omega}_{n,i,j} - \Phi^{\rm ex}_{n,i,j}\bigr|^2}
{\displaystyle
\sum_{(i,j)\in\mathcal I_S}
V_{i,j}\,
}
\right)^{1/2},
\label{eq:relL2}
\end{equation}
where 
$\phi^{\rm ex}_{n,i,j}$ denotes the exact solution evaluated consistently with
the discrete unknown e.g.\ at the cell centroid for cell averages. We report
\[
\|e\|_{2,\mathrm{reg}}^{\rm},\qquad
\|e\|_{2,\mathrm{cut}}^{\rm},\qquad
\|e\|_{2,\mathrm{all}}^{\rm}.
\]

The empirical convergence order is estimated between two successive grid
resolutions $h_i$ and $h_{i+1}$ using
\begin{equation}
p_{2,S} =
\frac{\log\!\left(\|e\|_{2,S}^{(i)} / \|e\|_{2,S}^{(i+1)}\right)}
     {\log\!\left(h_i / h_{i+1}\right)},
\qquad
S \in \{\mathrm{reg},\mathrm{cut},\mathrm{all}\}.
\label{eq:order}
\end{equation}

Several benchmarks are formulated in a single moving domain (one material
region), i.e.\ only one phase is physically present. In our two-fluid
framework, this is recovered by selecting a single active phase (say
$\Omega^-(t)$) and treating the complement as inactive.  The interface
$\Gamma(t)$ then acts as a prescribed moving boundary of $\Omega^-(t)$ and the
discrete update reduces to the single-phase counterpart of
\eqref{eq:st-bulk-divergence}-\eqref{eq:st-constitutive} without interfacial
coupling unknowns.  Concretely, the continuity/Henry closure and the
space-time interface balance are dropped and boundary conditions are imposed
directly on $\Gamma(t)$ (Dirichlet/Neumann/Robin), while the space-time
sweeping term and the geometric conservation law remain active and ensure exact
preservation during the motion.

\subsection{4D geometric integration: validation of the space-time moment integration engine}
\label{sec:val_4d_geom}

A central ingredient of the space-time formulation is the accurate evaluation
of time-integrated geometric moments over a slab
$[t_n,t_{n+1}]$. In three spatial dimensions, these moments correspond to
four-dimensional measures in $(x,y,z,t)$ and their computation requires
robust clipping and integration of $(d\!+\!1)$-dimensional polytopes.  To
validate this component independently of the PDE discretization,
we perform a purely geometric test in $\mathbb{R}^4$.

We consider a uniform Cartesian grid on a fixed four-dimensional box
$\Omega_4\subset\mathbb{R}^4$ with $N$ cells per direction and mesh size
$h=1/N$. For each test geometry, defined by a smooth level-set
$\Psi(\mathbf{x})$ in $\mathbb{R}^4$, we compute the hypervolume of the
restricted set $\{\Psi<0\}\cap\Omega_4$. The geometric quantities are evaluated using our
VOFI-based \cite{chierici_optimized_2022} integration algorithm extended to four dimensions (recursive reconstruction and numerical integration over the resulting 4-polytopes).

Table~\ref{tab:4d_volume_integration} reports the convergence of the
hypervolume integral for three representative 4D geometries: a
hypersphere, a hyperellipsoid and a sinusoidal slab. The exact reference
volumes are available in closed form for the hypersphere and the ellipsoid and
for the sinusoidal slab they are obtained from its analytic definition; the
corresponding values are listed in the table.

\begin{table}[h!]
\centering

%------------------ top row ------------------%
\begin{subtable}[t]{0.48\textwidth}
\centering
\setlength{\tabcolsep}{6pt}
\renewcommand{\arraystretch}{1.15}
\begin{tabular}{c c | c | c}
\hline
$N$ & $h$ & $V_{\rm num}$ & rel.\ err. \\
\hline\hline
4  & 0.25      & $7.4053150\mathrm{e}{-2}$ & $3.7055\mathrm{e}{-6}$  \\
6  & 0.166666  & $7.4052876\mathrm{e}{-2}$ & $3.0536\mathrm{e}{-9}$  \\
8  & 0.125     & $7.4052875\mathrm{e}{-2}$ & $1.7002\mathrm{e}{-9}$  \\
10 & 0.10      & $7.4052876\mathrm{e}{-2}$ & $1.1917\mathrm{e}{-10}$ \\
12 & 0.083333  & $7.4052876\mathrm{e}{-2}$ & $4.3828\mathrm{e}{-10}$ \\
16 & 0.0625    & $7.4052876\mathrm{e}{-2}$ & $4.5370\mathrm{e}{-13}$ \\
32 & 0.03125   & $7.4052876\mathrm{e}{-2}$ & $3.0266\mathrm{e}{-13}$ \\
\hline
\end{tabular}
\caption{\textbf{Hypersphere}.}
\label{tab:4d_vol_hypersphere}
\end{subtable}
\hfill
\begin{subtable}[t]{0.48\textwidth}
\centering
\setlength{\tabcolsep}{6pt}
\renewcommand{\arraystretch}{1.15}
\begin{tabular}{c c | c | c}
\hline
$N$ & $h$ & $V_{\rm num}$ & rel.\ err. \\
\hline\hline
4  & 0.25      & $8.3275552\mathrm{e}{-2}$ & $9.1814\mathrm{e}{-6}$  \\
6  & 0.166666  & $8.3274857\mathrm{e}{-2}$ & $8.3516\mathrm{e}{-7}$  \\
8  & 0.125     & $8.3274787\mathrm{e}{-2}$ & $8.6144\mathrm{e}{-10}$ \\
10 & 0.10      & $8.3274787\mathrm{e}{-2}$ & $2.2451\mathrm{e}{-10}$ \\
12 & 0.083333  & $8.3274787\mathrm{e}{-2}$ & $9.1379\mathrm{e}{-11}$ \\
16 & 0.0625    & $8.3274787\mathrm{e}{-2}$ & $7.9042\mathrm{e}{-11}$ \\
32 & 0.03125   & $8.3274787\mathrm{e}{-2}$ & $6.6589\mathrm{e}{-10}$ \\
\hline
\end{tabular}
\caption{\textbf{Hyperellipsoid}.}
\label{tab:4d_vol_hyperellipsoid}
\end{subtable}

\vspace{0.8em}

%------------------ bottom row ------------------%
\begin{subtable}[t]{0.62\textwidth}
\centering
\setlength{\tabcolsep}{6pt}
\renewcommand{\arraystretch}{1.15}
\begin{tabular}{c c | c | c}
\hline
$N$ & $h$ & $V_{\rm num}$ & rel.\ err. \\
\hline\hline
4  & 0.25      & $6.0082800\mathrm{e}{-1}$ & $1.3800\mathrm{e}{-3}$  \\
6  & 0.166666  & $5.9999988\mathrm{e}{-1}$ & $2.0558\mathrm{e}{-7}$  \\
8  & 0.125     & $6.0000000\mathrm{e}{-1}$ & $1.4730\mathrm{e}{-9}$  \\
10 & 0.10      & $6.0000000\mathrm{e}{-1}$ & $1.1591\mathrm{e}{-11}$ \\
12 & 0.083333  & $6.0000000\mathrm{e}{-1}$ & $4.0708\mathrm{e}{-14}$ \\
16 & 0.0625    & $6.0000000\mathrm{e}{-1}$ & $1.6098\mathrm{e}{-14}$ \\
32 & 0.03125   & $6.0000000\mathrm{e}{-1}$ & $7.0684\mathrm{e}{-14}$ \\
\hline
\end{tabular}
\caption{\textbf{Sinusoidal slab}.}
\label{tab:4d_vol_sinuslab}
\end{subtable}

\caption{Four-dimensional hypervolume integration for representative geometries.
$V_{\rm num}$ is computed by the 4D geometric integration engine and
``rel.\ err.'' denotes $|V_{\rm num}-V_{\rm ex}|/|V_{\rm ex}|$.
Exact values: hypersphere $V_{\rm ex}=0.07405287552192358$, hyperellipsoid
$V_{\rm ex}=0.08327478713419147$, sinusoidal slab $V_{\rm ex}=0.6$.}
\label{tab:4d_volume_integration}
\end{table}

Across the curved geometries (hypersphere and ellipsoid), the hypervolume error
drops by several orders of magnitude between coarse and moderate resolutions
and reaches near round-off levels on the finest grids (e.g.\ absolute errors of
$\mathcal{O}(10^{-14})$ for the hypersphere at $N\ge 16$). In this regime,
observed orders become unreliable and may appear non-monotone because the error
is dominated by floating-point effects and by the tolerance of the geometric
predicates rather than by the mesh size. The sinusoidal slab exhibits the same
behaviour on most grids (machine-level accuracy at $N=8,12,16$), while a few
intermediate resolutions show larger deviations; nevertheless, refinement
recovers the expected high accuracy. Overall, these results validate the
correctness and robustness of the four-dimensional clipping and integration
pipeline that underpins the space-time moments used by the moving-domain
operators.

\subsection{2D diffusion with an oscillating circular boundary}

We consider a transient diffusion problem in a circular domain whose boundary
oscillates periodically in time. The background domain is the fixed box
$\Omega_{\rm box}=[0,4]^2$ and the physical domain is
\[
\Omega(t) = \{(x,y)\in\Omega_{\rm box} : r(x,y) < R(t)\},
\qquad
r(x,y)=\|(x,y)-\mathbf{x}_c\|,
\qquad
\mathbf{x}_c=(2,2).
\]
The radius follows the smooth motion
\begin{equation}
R(t)=R_{\mathrm{mean}}+R_{\mathrm{amp}}\sin\!\left(2\pi \frac{t}{T}\right),
\label{eq:oscillating-radius}
\end{equation}
with $R_{\mathrm{mean}}=1.0$, $R_{\mathrm{amp}}=0.5$ and $T=1.0$. The moving
boundary $\Gamma(t)=\partial\Omega(t)$ therefore sweeps through the Cartesian
mesh and continuously generates fresh and dead cut cells over the simulation.

Inside $\Omega(t)$ we solve the diffusion equation
\begin{equation}
\partial_t \phi = D\,\Delta \phi + f(x,y,t),
\qquad (x,y)\in\Omega(t),
\label{eq:oscillating-disk-pde}
\end{equation}
with constant diffusivity $D$ and Dirichlet boundary conditions prescribed on the moving
boundary,
\begin{equation}
\phi(x,y,t)=\phi_{\mathrm{ex}}(x,y,t)
\qquad\text{on }\Gamma(t).
\label{eq:oscillating-disk-bc}
\end{equation}
The pair $(\phi_{\mathrm{ex}},f)$ is manufactured so that $\phi_{\mathrm{ex}}$
satisfies \eqref{eq:oscillating-disk-pde}-\eqref{eq:oscillating-disk-bc}
exactly in the evolving domain; we use
\begin{equation}
\phi_{\mathrm{ex}}(x,y,t)=R(t)\,\cos(\pi x)\,\cos(\pi y),
\label{eq:oscillating-disk-exact}
\end{equation}
which yields the forcing term
\begin{equation}
f(x,y,t)=\cos(\pi x)\,\cos(\pi y)\left[
\dot R(t)+2\pi^2 D\,R(t)
\right],
\qquad
\dot R(t)=\frac{2\pi R_{\mathrm{amp}}}{T}\cos\!\left(2\pi\frac{t}{T}\right).
\label{eq:oscillating-disk-forcing}
\end{equation}

We take $D=0.1$, integrate to $t_f=T$ with $\Delta t= 0.25\min(\Delta x, \Delta y)$ using a $\theta=1/2$ midpoint scheme. Errors are evaluated at the final time $t_f$ using cut-cell weighted $L^2$ norms
over regular cells, cut cells and all cells. Table~\ref{tab:2D-moving-osc}
reports the corresponding errors and observed convergence rates.
\begin{table}[h!]
\centering
\setlength{\tabcolsep}{6pt}
\renewcommand{\arraystretch}{1.2}
\begin{tabular}{c c | c c c | c c c}
\hline
$h$ & $N_{\rm diam}$ &
$\|e(t_f)\|_{2,\mathrm{reg}}$ &
$\|e(t_f)\|_{2,\mathrm{cut}}$ &
$\|e(t_f)\|_{2,\mathrm{all}}$ &
$p_{\mathrm{reg}}$ & $p_{\mathrm{cut}}$ & $p_{\mathrm{all}}$ \\
\hline
1.0     & 3  & $2.315\mathrm{e}{-1}$ & $1.598\mathrm{e}{0}$  & $1.615\mathrm{e}{0}$   & -    & -    & -    \\
0.5     & 6  & $1.496\mathrm{e}{-1}$ & $1.600\mathrm{e}{-1}$ & $2.191\mathrm{e}{-1}$  & 0.63 & 3.32 & 2.88 \\
0.25    & 12 & $7.056\mathrm{e}{-2}$ & $3.532\mathrm{e}{-2}$ & $7.890\mathrm{e}{-2}$  & 1.08 & 2.18 & 1.47 \\
0.125   & 24 & $2.780\mathrm{e}{-2}$ & $8.013\mathrm{e}{-3}$ & $2.893\mathrm{e}{-2}$  & 1.34 & 2.14 & 1.45 \\
0.0625  & 42 & $1.210\mathrm{e}{-2}$ & $2.565\mathrm{e}{-3}$ & $1.237\mathrm{e}{-2}$  & 1.20 & 1.64 & 1.23 \\
0.03125 & 84 & $5.368\mathrm{e}{-3}$ & $6.835\mathrm{e}{-4}$ & $5.411\mathrm{e}{-3}$  & 1.17 & 1.91 & 1.19 \\
\hline
fit     & -  & - & - & - & 1.21 & 1.95 & 1.34 \\
\hline
\end{tabular}
\caption{$L^2$ error at final time $t_f$ for the oscillating-disk diffusion test.}
\label{tab:2D-moving-osc}
\end{table}

Table~\ref{tab:2D-moving-osc} demonstrates a systematic reduction of the error
under mesh refinement, on regular cells, on cut cells and on the full set of
cells. Remarkably, the method remains robust even in the most under-resolved
configuration ($N_{\rm diam}=3$), where the moving boundary sweeps through only
a handful of cells and repeatedly generates fresh and dead cut cells within a
single period. As expected, this regime produces large cut-cell errors but it
constitutes a stringent stress test of the space-time balance in the presence
of very small and rapidly evolving phase-restricted volumes.

Once the interface is better resolved ($h\le 0.5$), the cut-cell contribution
decreases rapidly and exhibits super-linear convergence, with an overall rate
close to second order. On the finest grids, the cut-cell error becomes smaller
than the regular-cell error, indicating that the space-time sweeping treatment
and the fresh/dead-cell closure do not induce a degradation of accuracy near
the moving boundary.

\begin{figure}[h]
\centering
\includegraphics[width=\linewidth]{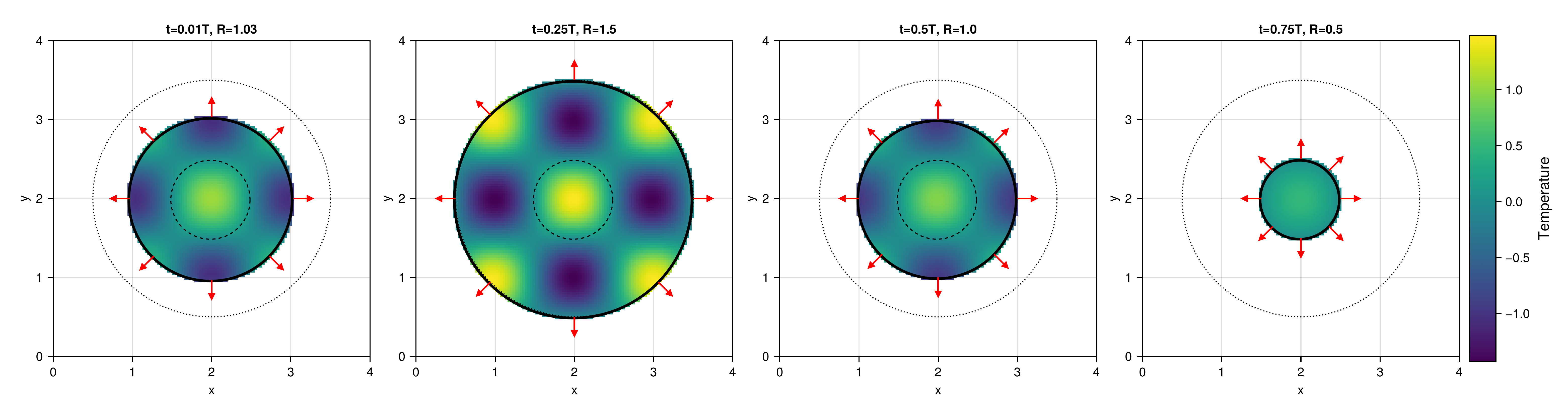}
\caption{Snapshots at four phases of one oscillation period
($t=0$, $T/4$, $T/2$, $3T/4$). The solid curve shows the exact moving boundary
$r=R(t)$ and the colormap shows the numerical solution $\phi$.}
\label{fig:oscillating_disk_snapshots}
\end{figure}

Figure~\ref{fig:oscillating_disk_snapshots} further confirms that the numerical
solution remains smooth and bounded throughout the oscillation cycle, with no
visible artifacts correlated with the creation or disappearance of cut cells.
The solution matches the prescribed Dirichlet boundary condition along $\Gamma(t)$ at all
phases, while the interior field evolves continuously during expansion and
contraction, illustrating the robustness of the space-time flux balance across
topology changes.

\subsection{2D Diffusion with multiple bodies (McCorquodale-Colella)}
\label{sec:jc_moving_ellipses}

We consider the two-dimensional moving-interface benchmark of McCorquodale and
Colella \cite{mccorquodale_cartesian_2001}, in which several rigid ellipses translate with prescribed velocities
inside a fixed Cartesian box. The background domain is the fixed Cartesian box
\(
\Omega_{\rm box}=[-1.5,\,1.5]\times[-1.0,\,1.0],
\)
and the time-dependent physical domain is obtained by removing the interiors of
three moving ellipses,
\[
\Omega(t)=\Omega_{\rm box}\setminus\bigcup_{k=1}^3 E_k(t),
\qquad
\Gamma(t)=\partial\Omega(t)
=
\partial\Omega_{\rm box}\ \cup\ \bigcup_{k=1}^3 \Gamma_k(t),
\]
where $\Gamma_k(t)=\partial E_k(t)$ denotes the boundary of ellipse $k$.  As the
ellipses translate across the mesh, the interface $\Gamma(t)$ sweeps through
cells and repeatedly creates fresh and dead cut cells.

Let $\gamma=\sqrt{2}/15$. At $t=0$, ellipse $k$ is defined by its center
$(x_k^0,y_k^0)$ and semi-axes $(a_k,b_k)$,
\begin{align*}
(x_1^0,y_1^0,a_1,b_1)&=(-6\gamma,-5\gamma,3\gamma,2\gamma),\\
(x_2^0,y_2^0,a_2,b_2)&=(10\gamma,-7\gamma,2\gamma,\gamma),\\
(x_3^0,y_3^0,a_3,b_3)&=(7\gamma,3\gamma,1.5\gamma,2\gamma).
\end{align*}
and translated with constant velocities
\[
(u_1,v_1)=(-0.10,0.20),\qquad
(u_2,v_2)=(-0.15,0.15),\qquad
(u_3,v_3)=(-0.20,0.20).
\]
Thus, the ellipse centers evolve as
\[
x_k(t)=x_k^0+u_k t,\qquad y_k(t)=y_k^0+v_k t,
\]
and the moving boundaries are
\[
\Gamma_k(t)=\left\{(x,y):\ 
\left(\frac{x-x_k(t)}{a_k}\right)^2
+
\left(\frac{y-y_k(t)}{b_k}\right)^2
=1\right\}.
\]

On $\Omega(t)$ we solve the transient diffusion problem
\begin{equation}
\partial_t \phi = \Delta \phi + f(x,y,t),
\qquad (x,y)\in\Omega(t),
\label{eq:jc_pde}
\end{equation}
with Dirichlet boundary conditions prescribed on the entire boundary $\Gamma(t)$,
\begin{equation}
\phi(x,y,t)=\phi_{\rm ex}(x,y,t)\qquad \text{on }\Gamma(t).
\label{eq:jc_bc}
\end{equation}
The manufactured reference field and corresponding forcing are chosen as
\begin{equation}
\phi_{\rm ex}(x,y,t)
=
\frac{4}{5\pi\,(t+1)}
\exp\!\left(-\frac{x^2+y^2}{5(t+1)}\right),
\label{eq:jc_exact}
\end{equation}
\begin{equation}
f(x,y,t)
=
\frac{4\,(x^{2}+y^{2}-5(t+1))}
     {125\,\pi\,(t+1)^{3}}
\exp\!\left(-\frac{x^{2}+y^{2}}{5(t+1)}\right),
\label{eq:jc_forcing}
\end{equation}
so that \eqref{eq:jc_exact} satisfies \eqref{eq:jc_pde}-\eqref{eq:jc_bc} exactly
in the moving domain.

Table~\ref{tab:JC-moving-dirichlet} reports the cut-cell weighted $L^2$ error at
final time $t_f$, split into regular cells, cut cells and all cells. Here
$N_{\rm diam}$ denotes the number of cells per characteristic domain diameter.

\begin{table}[h!]
\centering
\setlength{\tabcolsep}{6pt}
\renewcommand{\arraystretch}{1.2}
\begin{tabular}{c c | c c c | c c c}
\hline
$h$ & $N_{\rm diam}$ &
$\|e(t_f)\|_{2,\mathrm{reg}}$ &
$\|e(t_f)\|_{2,\mathrm{cut}}$ &
$\|e(t_f)\|_{2,\mathrm{all}}$ &
$p_{\mathrm{reg}}$ & $p_{\mathrm{cut}}$ & $p_{\mathrm{all}}$ \\
\hline
0.50        & 1  & -              & -              & -              & -    & -    & -    \\
0.22        & 2  & -              & $4.509\mathrm{e}{-3}$  & $4.509\mathrm{e}{-3}$  & -    & -    & -    \\
0.125       & 4  & $8.410\mathrm{e}{-4}$  & $1.786\mathrm{e}{-3}$  & $1.974\mathrm{e}{-3}$  & -    & 1.61 & 1.44 \\
0.0606      & 7  & $4.053\mathrm{e}{-4}$  & $3.902\mathrm{e}{-4}$  & $5.626\mathrm{e}{-4}$  & 1.01 & 2.10 & 1.73 \\
0.0317      & 12 & $1.146\mathrm{e}{-4}$  & $1.108\mathrm{e}{-4}$  & $1.594\mathrm{e}{-4}$  & 1.95 & 1.95 & 1.95 \\
0.0157      & 24 & $3.620\mathrm{e}{-5}$  & $3.380\mathrm{e}{-5}$  & $4.953\mathrm{e}{-5}$  & 1.64 & 1.69 & 1.67 \\
\hline
fit & - & - & - & - & 1.55 & 1.89 & 1.74 \\
\hline
\end{tabular}
\caption{$L_2$ error at final time $t_f$ for the McCorquodale-Colella moving-ellipses
Dirichlet test.}
\label{tab:JC-moving-dirichlet}
\end{table}

\begin{figure}[h]
\centering
\includegraphics[width=0.75\linewidth]{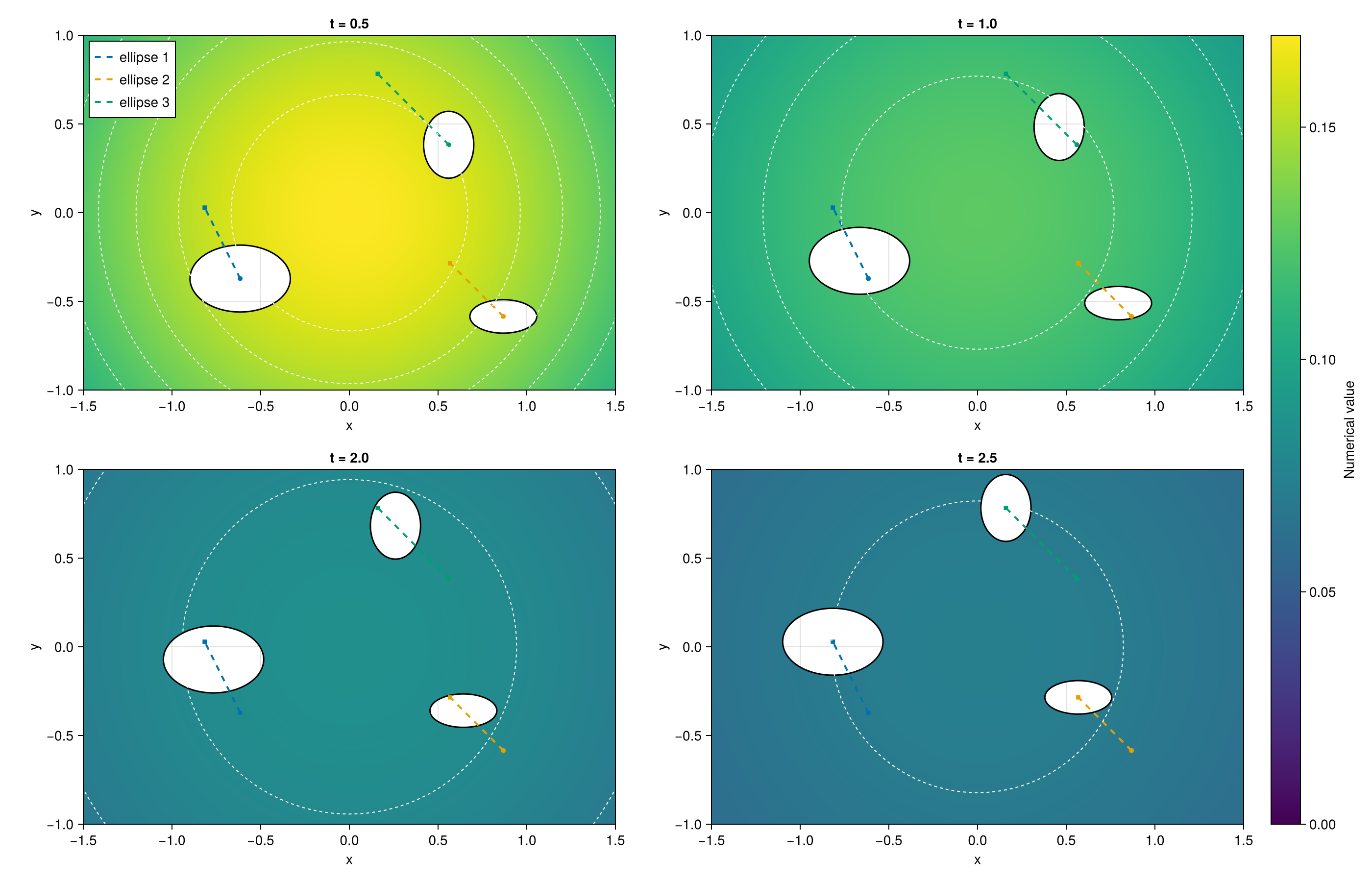}
\caption{Snapshots of the moving-ellipses test at four instants
($t=0.5$, $t=1.0$, $t=2.0$, $t=2.5$). The solid curves show the translated
elliptic boundaries $\Gamma_k(t)$ and the colormap shows the numerical solution
$\phi$.}
\label{fig:jc_moving_ellipses_snapshots}
\end{figure}

The two coarsest meshes are severely under-resolved (few cells per ellipse),
yet the method remains robust and produces a bounded solution while the three
bodies translate and continuously trigger fresh/dead cut-cell events.
From $h\le 0.125$ onward, the errors decrease systematically on both regular
and cut-cell subsets. The cut-cell error exhibits a clear super-linear regime
and approaches second-order behaviour under refinement, leading to an overall
least-squares rate close to two on the finest levels (Table~\ref{tab:JC-moving-dirichlet}).
The comparable decay of $\|e\|_{2,\mathrm{reg}}$ and $\|e\|_{2,\mathrm{cut}}$
indicates that the space-time sweeping treatment and the moving Dirichlet
enforcement do not degrade accuracy near the translating boundaries. This test
also exercises the simultaneous handling of multiple embedded bodies:
three distinct ellipses translate concurrently and contribute to the global set
of cut cells. The solver remains robust in this multi-body configuration.
Figure~\ref{fig:jc_moving_ellipses_snapshots} shows smooth snapshots across the
trajectory, with no visible artifacts correlated with the repeated creation and
removal of cut cells.

\subsection{3D Schwartz-Colella diffusion on an expanding sphere}
\label{sec:sc3d_expanding_sphere}

We reproduce the three-dimensional moving-boundary benchmark of
Schwartz and Colella \cite{schwartz_cartesian_2006}, in which diffusion is solved in a domain bounded by a
prescribed expanding sphere. The background domain is the fixed Cartesian box
$\Omega_{\rm box}=[-1,1]^3$, while the physical domain is the time-dependent ball
\[
\Omega(t)=\{(x,y,z)\in\Omega_{\rm box}:\ r(x,y,z)<R(t)\},
\qquad
r(x,y,z)=\|(x,y,z)\|,
\qquad
R(t)=0.392+t,
\]
with moving boundary $\Gamma(t)=\partial\Omega(t)=\{r=R(t)\}$. The interface
motion sweeps through the mesh and continuously creates fresh and dead cut
cells over the simulation interval.

Inside $\Omega(t)$ we solve the transient diffusion equation
\begin{equation}
\partial_t \phi = \Delta \phi + f(x,y,z,t),
\qquad (x,y,z)\in\Omega(t),
\label{eq:sc3d_pde}
\end{equation}
with Dirichlet boundary conditions prescribed on the immersed boundary,
\begin{equation}
\phi(x,y,z,t)=\phi_{\rm ex}(x,y,z,t)
\qquad\text{on }\Gamma(t).
\label{eq:sc3d_bc}
\end{equation}
The manufactured reference field is chosen as
\begin{equation}
\phi_{\rm ex}(x,y,z,t)
=
\frac{4}{5\pi\,(t+1)}
\exp\!\left(-\frac{x^2+y^2+z^2}{5(t+1)}\right),
\label{eq:sc3d_exact}
\end{equation}
and the forcing term is
\begin{equation}
f(x,y,z,t)
=
\frac{4\,\bigl(x^2+y^2+z^2+5(t+1)\bigr)}
     {125\,\pi\,(t+1)^3}
\exp\!\left(-\frac{x^2+y^2+z^2}{5(t+1)}\right),
\label{eq:sc3d_forcing}
\end{equation}
so that \eqref{eq:sc3d_exact} satisfies \eqref{eq:sc3d_pde}-\eqref{eq:sc3d_bc}
exactly in the evolving domain.

Errors are evaluated at the final time $t_f$ using cut-cell weighted $L^2$ norms
over regular cells, cut cells and all active cells. Table~\ref{tab:sc3d_exp_sphere}
reports the corresponding errors and pairwise observed orders under refinement.

\begin{table}[h!]
\centering
\setlength{\tabcolsep}{6pt}
\renewcommand{\arraystretch}{1.15}
\begin{tabular}{c c | c c c | c c c}
\hline
$h$ & $N_{in}$ & 
$\|e(t_f)\|_{2,\mathrm{all}}$ &
$\|e(t_f)\|_{2,\mathrm{reg}}$ &
$\|e(t_f)\|_{2,\mathrm{cut}}$ &
$p_{\mathrm{all}}$ & $p_{\mathrm{reg}}$ & $p_{\mathrm{cut}}$ \\
\hline
0.25      & 18   & $1.4110\mathrm{e}{-2}$ & $7.5513\mathrm{e}{-3}$ & $1.19198\mathrm{e}{-2}$ &  -    &  -    &  -    \\
0.125     & 147  & $6.9784\mathrm{e}{-3}$ & $5.8039\mathrm{e}{-3}$ & $3.8745\mathrm{e}{-3}$  & 1.016 & 0.380 & 1.621 \\
0.083333  & 611  & $4.0096\mathrm{e}{-3}$ & $3.6258\mathrm{e}{-3}$ & $1.7121\mathrm{e}{-3}$  & 1.367 & 1.160 & 2.014 \\
0.0625    & 1479 & $2.7139\mathrm{e}{-3}$ & $2.5211\mathrm{e}{-3}$ & $1.0047\mathrm{e}{-3}$  & 1.357 & 1.263 & 1.853 \\
\hline
fit  & - & - & - & - & 1.36 & 1.20 & 1.95 \\
\hline
\end{tabular}
\caption{$L^2$ error at final time $t_f$ for the 3D Schwartz--Colella expanding-sphere test.
Errors are split into contributions from all active cells, regular cells and cut cells.
Here $N_{in}$ denotes the number of active cells inside the sphere.}
\label{tab:sc3d_exp_sphere}
\end{table}

The method remains stable while the sphere expands and sweeps across the
Cartesian mesh, continuously creating fresh cut cells near $\Gamma(t)$.
Table~\ref{tab:sc3d_exp_sphere} shows consistent error reduction under refinement
for all three subsets. The global error $\|e\|_{2,\mathrm{all}}$ exhibits
super-linear convergence. Cut-cell errors converge even faster (orders $\approx 1.6$-$2.0$),
indicating that the space-time sweeping treatment and the fresh/dead-cell
closure do not degrade accuracy near the moving boundary. The regular-cell
contribution converges more slowly on the coarsest refinement but approaches a
super-linear regime as the sphere becomes better resolved. 

\subsection{Manufactured solution with an oscillating two-phase interface}
\label{nummotion:manufactured_two_phase}

To assess the accuracy and robustness of the moving two-phase solver, we
design a manufactured configuration with a time-periodic vertical sharp
interface. The computational box is $\Omega=[0,4]^2$ and the interface location
is prescribed as
\begin{equation}
s(t)=s_0 + A \sin(\omega t),
\label{eq:2ph_s_of_t}
\end{equation}
so that the moving interface is $\Gamma(t)=\{x=s(t)\}$ and the two phases are
\[
\Omega^-(t)=\{(x,y)\in\Omega:\ x<s(t)\},
\qquad
\Omega^+(t)=\{(x,y)\in\Omega:\ x>s(t)\}.
\]
We use the unit normal $\mathbf n=(1,0)$ pointing from $\Omega^-(t)$ to
$\Omega^+(t)$, hence $\partial_n=\partial_x$ and the prescribed normal
velocity is
\begin{equation}
w(t)=\dot s(t)=A\omega\cos(\omega t).
\label{eq:2ph_w_of_t}
\end{equation}

In each phase we solve the transient diffusion model
\begin{equation}
c_p^\pm\,\partial_t\phi^\pm
=
\nabla\!\cdot\!\bigl(D^\pm\nabla\phi^\pm\bigr)
+ f^\pm(x,y,t),
\qquad (x,y)\in\Omega^\pm(t),
\label{eq:2ph_mms_pde}
\end{equation}
with Dirichlet boundary conditions on the outer boundary.
Across $\Gamma(t)$ we impose the two-fluid coupling conditions
\begin{equation}
\llbracket \phi \rrbracket = 0,
\qquad
\llbracket D\,\partial_n\phi \rrbracket = 0,
\qquad \text{on }\Gamma(t),
\label{eq:2ph_mms_jumps}
\end{equation}
corresponding to continuity of temperature and diffusive heat flux, with no
interfacial storage or sources.

We choose phase fields of the form
\begin{equation}
\phi^\pm(x,y,t)
=
A^\pm\,\bigl(x-s(t)\bigr)\,x(4-x)\,y(4-y)\,e^{-t},
\label{eq:2ph_mms_phi}
\end{equation}
with the coefficients
\(
A^- = D^+,
A^+ = D^-.
\)
This construction enforces $\phi^\pm=0$ on $\partial\Omega$ and also $\llbracket \phi \rrbracket=0$.
Moreover,
\[
\partial_x\phi^\pm(s(t),y,t)
=
A^\pm\,s(t)\bigl(4-s(t)\bigr)\,y(4-y)\,e^{-t},
\]
so that the choice $A^-=D^+$ and $A^+=D^-$ yields
$D^-\partial_x\phi^-(s(t),y,t)=D^+\partial_x\phi^+(s(t),y,t)$, i.e.
$\llbracket D\,\partial_n\phi \rrbracket=0$.

The forcing terms are defined analytically by substitution of
\eqref{eq:2ph_mms_phi} into \eqref{eq:2ph_mms_pde},
\begin{equation}
f^\pm(x,y,t)
=
c_p^\pm\,\partial_t\phi^\pm
-
\nabla\!\cdot\!\bigl(D^\pm\nabla\phi^\pm\bigr),
\label{eq:2ph_mms_forcing}
\end{equation}
with $\partial_t\phi^\pm$ containing the interface-motion contribution through
$s'(t)=A\omega\cos(\omega t)$. In particular, increasing $\omega$ directly
amplifies the time-dependent part of the forcing and increases the sweeping
speed $|w(t)|$, thus intensifying the creation/removal of cut cells and making
the test more demanding for the space-time update.

In the results below we take
\(
c_p^\pm=1, D^- = 0.1, D^+=1.0,
s_0=2, A=1,
\)
integrate to a final time $t_f=0.5$ with a midpoint time scheme and measure cut-cell weighted $L^2$ errors at
$t_f$ over regular cells, cut cells and all active cells. As $\Gamma(t)$
oscillates, both phases repeatedly generate fresh and dead cut cells, providing
a direct validation of the space-time two-phase coupling in the presence of
topology changes.

\begin{figure}[h]
\centering
\includegraphics[width=0.75\linewidth]{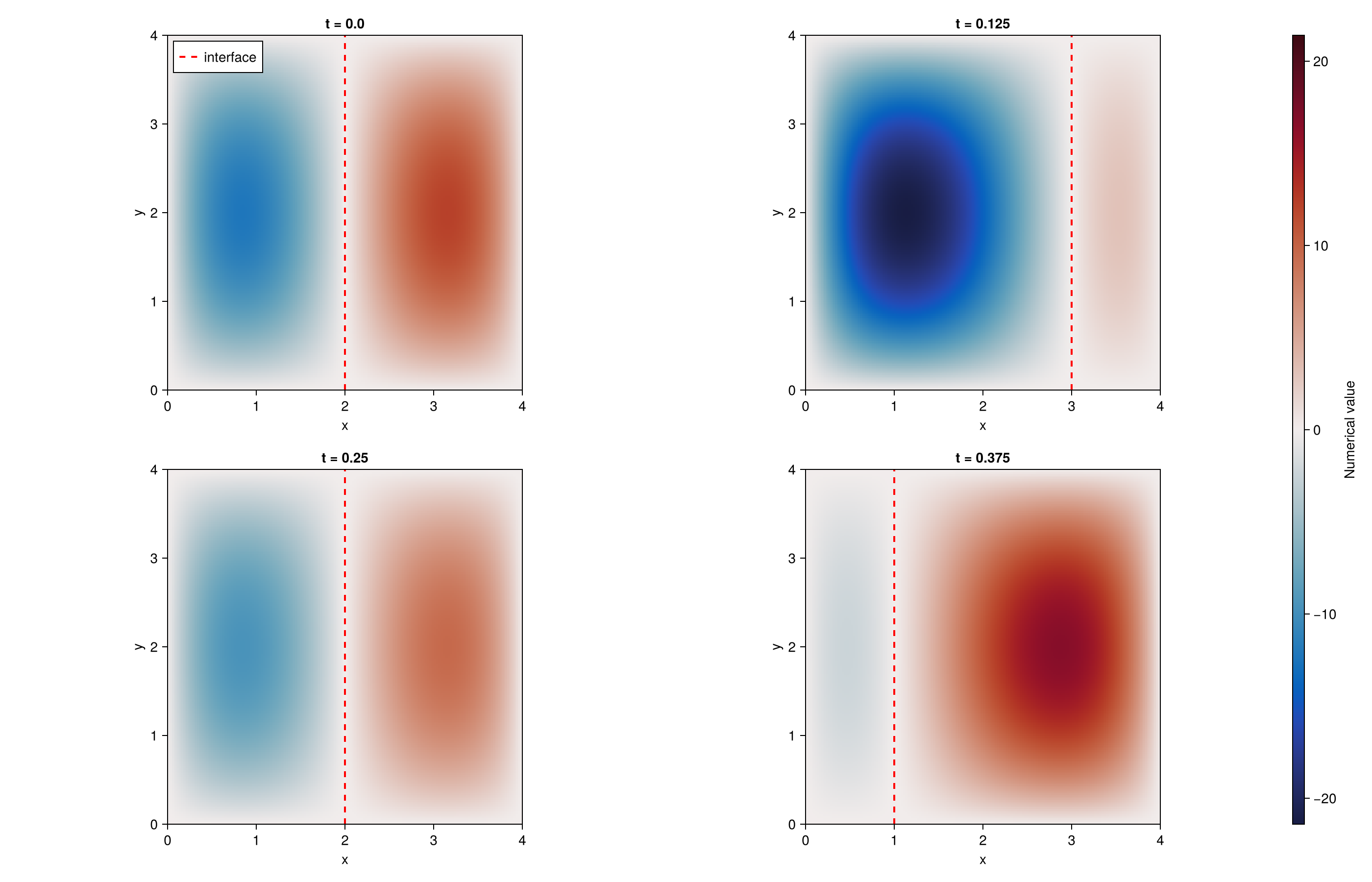}
\caption{Snapshots of the moving vertical interface test at four instants
($t=0$, $t=0.125$, $t=0.25$, $t=0.375$). The red dashed line shows the prescribed moving
boundary and the colormap represents $\phi$ on planar slices.}
\label{fig:mms_osc_2ph_snapshots}
\end{figure}

Table~\ref{tab:mms-osc-2ph} reports the errors and pairwise observed orders on a
sequence of $N\times N$ grids.

\begin{table}[h!]
\centering
\setlength{\tabcolsep}{6pt}
\renewcommand{\arraystretch}{1.2}
\begin{tabular}{c | c c c | c c c}
\hline
$h$ &
$\|e(t_f)\|_{2,\mathrm{reg}}$ &
$\|e(t_f)\|_{2,\mathrm{cut}}$ &
$\|e(t_f)\|_{2,\mathrm{all}}$ &
$p_{\mathrm{reg}}$ &
$p_{\mathrm{cut}}$ &
$p_{\mathrm{all}}$ \\
\hline
1.0     & $2.3827\mathrm{e}{-1}$ & $3.0388\mathrm{e}{-1}$ & $3.8616\mathrm{e}{-1}$ &  -    &  -    &  -    \\
0.5     & $1.9904\mathrm{e}{-1}$ & $1.8845\mathrm{e}{-1}$ & $2.5577\mathrm{e}{-1}$ & 0.26  & 0.69  & 0.59  \\
0.25    & $8.4559\mathrm{e}{-2}$ & $5.9272\mathrm{e}{-2}$ & $9.1495\mathrm{e}{-2}$ & 1.24  & 1.67  & 1.48  \\
0.125   & $2.8215\mathrm{e}{-2}$ & $1.6183\mathrm{e}{-2}$ & $3.0286\mathrm{e}{-2}$ & 1.58  & 1.87  & 1.60  \\
0.0625  & $8.0159\mathrm{e}{-3}$ & $2.3389\mathrm{e}{-3}$ & $8.2765\mathrm{e}{-3}$ & 1.82  & 2.79  & 1.87  \\
0.03125 & $2.1327\mathrm{e}{-3}$ & $4.5489\mathrm{e}{-4}$ & $2.1614\mathrm{e}{-3}$ & 1.91  & 2.36  & 1.94  \\
\hline
fit  & - & - & - & 1.86 & 2.58 & 1.90 \\
\hline
\end{tabular}
\caption{$L^2$ error and convergence rates for the manufactured oscillating two-phase
vertical-interface test.}
\label{tab:mms-osc-2ph}
\end{table}

The global error decreases monotonically with refinement and enters an
approximately second-order regime on the finest levels. The cut-cell error
converges faster than the regular-cell contribution on refined grids, showing
that the space-time sweeping treatment and the two-phase coupling across
$\Gamma(t)$ remain accurate despite repeated creation and removal of
phase-restricted volumes.

To probe robustness with respect to interface speed, we repeat the same
manufactured two-phase test while increasing the oscillation frequency
$\omega$ in $s(t)=s_0 + A\sin(\omega t)$ (and thus the normal velocity
$w(t)=A\omega\cos(\omega t)$). Increasing $\omega$ directly amplifies the
time-dependent component of the forcing through $s'(t)$ and increases the
sweeping speed $|w(t)|$, thereby intensifying the creation/removal of cut cells
within each time slab.

Tables~\ref{tab:2ph_freq_fit}-\ref{tab:2ph_freq_levels} summarize the results.
For moderate frequencies ($\omega\le 8\pi$), the global error
$\|e(t_f)\|_{2,\mathrm{all}}$ decreases monotonically with refinement and enters
a near second-order regime on the finest grids, with least-squares rates
$p_{\mathrm{all}}\approx 1.6$-$1.9$.
The cut-cell contribution remains superlinear,
showing that the space-time sweeping treatment and two-fluid coupling across
$\Gamma(t)$ do not degrade accuracy at the moving interface despite repeated
fresh/dead events. As $\omega$ increases further, the fitted global rate decreases (e.g.\ around
$p_{\mathrm{all}}\approx 1.2$ at $\omega=16\pi$ on the same fine-grid window),
which is consistent with the test becoming more demanding due to faster
geometry motion and stronger time-dependent forcing.
At $\omega=32\pi$, the error is not monotone across the three finest grids,
so a single asymptotic spatial rate is not meaningful; nonetheless, the solution
remains stable and bounded and the finest-grid errors remain comparable to the
lower-frequency cases.

\begin{table}[h!]
\centering
\setlength{\tabcolsep}{6pt}
\renewcommand{\arraystretch}{1.15}
\begin{tabular}{c | c c c | c c c}
\hline
$\omega$ &
$\|e(t_f)\|_{2,\mathrm{all}}(h_{\min})$ &
$\|e(t_f)\|_{2,\mathrm{reg}}(h_{\min})$ &
$\|e(t_f)\|_{2,\mathrm{cut}}(h_{\min})$ &
$p_{\mathrm{all}}$ &
$p_{\mathrm{reg}}$ &
$p_{\mathrm{cut}}$ \\
\hline
$2\pi$  & $2.1614\mathrm{e}{-3}$ & $2.1327\mathrm{e}{-3}$ & $4.5489\mathrm{e}{-4}$ & 1.90 & 1.86 & 2.58 \\
$4\pi$  & $2.1577\mathrm{e}{-3}$ & $2.1206\mathrm{e}{-3}$ & $3.9829\mathrm{e}{-4}$ & 1.83 & 1.82 & 2.46 \\
$8\pi$  & $1.9600\mathrm{e}{-3}$ & $1.9342\mathrm{e}{-3}$ & $4.1152\mathrm{e}{-4}$ & 1.62 & 1.59 & 2.26 \\
$16\pi$ & $1.6749\mathrm{e}{-3}$ & $1.6481\mathrm{e}{-3}$ & $2.9868\mathrm{e}{-4}$ & 1.19 & 1.16 & 1.86 \\
$32\pi$ & $9.0455\mathrm{e}{-4}$ & $8.9578\mathrm{e}{-4}$ & $1.6759\mathrm{e}{-4}$ & - & - & - \\
\hline
\end{tabular}
\caption{Frequency sweep for the manufactured oscillating two-phase interface.
Reported errors are at the finest mesh $h_{\min}=0.03125$.
Least-squares orders are fitted over $h\in\{0.125,\,0.0625,\,0.03125\}$.
For $\omega=32\pi$, the error is not monotone on this window, hence no single
fit is reported.}
\label{tab:2ph_freq_fit}
\end{table}

\begin{table}[h!]
\centering
\setlength{\tabcolsep}{6pt}
\renewcommand{\arraystretch}{1.15}
\begin{tabular}{c | c c c | c c c}
\hline
$\omega$ &
\multicolumn{3}{c|}{$h=0.125$} &
\multicolumn{3}{c}{$h=0.03125$} \\
 & $\|e\|_{2,\mathrm{all}}$ & $\|e\|_{2,\mathrm{reg}}$ & $\|e\|_{2,\mathrm{cut}}$
 & $\|e\|_{2,\mathrm{all}}$ & $\|e\|_{2,\mathrm{reg}}$ & $\|e\|_{2,\mathrm{cut}}$ \\
\hline
$2\pi$  & $3.0286\mathrm{e}{-2}$ & $2.8215\mathrm{e}{-2}$ & $1.6183\mathrm{e}{-2}$
       & $2.1614\mathrm{e}{-3}$ & $2.1327\mathrm{e}{-3}$ & $4.5489\mathrm{e}{-4}$ \\
$4\pi$  & $2.7262\mathrm{e}{-2}$ & $2.6297\mathrm{e}{-2}$ & $1.2065\mathrm{e}{-2}$
       & $2.1577\mathrm{e}{-3}$ & $2.1206\mathrm{e}{-3}$ & $3.9829\mathrm{e}{-4}$ \\
$8\pi$  & $1.8595\mathrm{e}{-2}$ & $1.7534\mathrm{e}{-2}$ & $9.4710\mathrm{e}{-3}$
       & $1.9600\mathrm{e}{-3}$ & $1.9342\mathrm{e}{-3}$ & $4.1152\mathrm{e}{-4}$ \\
$16\pi$ & $8.7483\mathrm{e}{-3}$ & $8.2674\mathrm{e}{-3}$ & $3.9314\mathrm{e}{-3}$
       & $1.6749\mathrm{e}{-3}$ & $1.6481\mathrm{e}{-3}$ & $2.9868\mathrm{e}{-4}$ \\
$32\pi$ & $6.0882\mathrm{e}{-4}$ & $4.3472\mathrm{e}{-4}$ & $4.2623\mathrm{e}{-4}$
       & $9.0455\mathrm{e}{-4}$ & $8.9578\mathrm{e}{-4}$ & $1.6759\mathrm{e}{-4}$ \\
\hline
\end{tabular}
\caption{Errors versus oscillation frequency at two representative mesh sizes.
The $\omega=32\pi$ case illustrates a non-monotone trend on fine grids, typical
of a very demanding sweeping regime.}
\label{tab:2ph_freq_levels}
\end{table}

\section{Conclusions and outlook}
\label{sec:conclusion}

This work extends our Cartesian cut-cell finite-volume framework to prescribed
moving geometries through a space-time formulation. Over each time slab
$[t_n,t_{n+1}]$, the evolving control volumes are treated as
$(d\!+\!1)$-dimensional space-time cut cells and all geometric moments
(volumes, apertures and auxiliary moments) are replaced by their time-integrated
counterparts. Starting from Reynolds’ transport theorem, we derived fully
conservative discrete balances in which the interface motion appears as an interface contribution that is closed by a geometric conservation law. This
construction preserves the static algebraic structure: the same local
divergence/gradient operators are reused, with only the geometric measures modified.

A first contribution is the geometric backbone required by the space-time
operators. We showed that the moment-based machinery lifts naturally to
higher-dimensional integration and validated a 4D volume/surface integration
engine on representative hypershapes (hypersphere, hyperellipsoid and a
sinusoidal slab). The results confirm that the extended VOFI-based integration
engine provides accurate hypervolume measures under refinement and is suitable
for assembling the time-integrated moments that drive the moving cut-cell scheme.

A second contribution is a robust treatment of topology changes within a time
step. Fresh and dead cut cells arise when the interface sweeps across
a fixed mesh. We introduced a state-dependent closure that keeps the
space-time constitutive fluxes well-defined even when one endpoint phase volume
vanishes, while maintaining conservation. Across all moving benchmarks,
no special-case remeshing or ad hoc mass fixes are required: the conservative
update follows directly from the space-time boundary fluxes.

The numerical validation suite supports these claims on increasingly demanding
moving configurations. In the monophasic oscillating-disk diffusion problem,
the method remains stable and bounded even in an under-resolved regime with only
a few points across the diameter, while exhibiting rapid error reduction once
the interface is better resolved. The McCorquodale-Colella moving-ellipses test
demonstrates that multiple translating bodies with different aspect ratios can
be handled simultaneously, recovering essentially second-order convergence when
the motion is smooth. In three dimensions, the Schwartz-Colella expanding-sphere
benchmark confirms that the space-time construction extends directly to $d=3$:
errors decrease systematically with refinement and the cut-cell contribution
converges at least as fast as the regular region, indicating that the
time-integrated geometric weights do not degrade accuracy near the moving
embedded boundary. Finally, the two-phase manufactured test validates the full
moving two-fluid coupling: continuity of the scalar and of the diffusive normal
flux is enforced across $\Gamma(t)$ while both phases experience repeated
fresh/dead events. The observed convergence approaches second order on refined
meshes and additional runs with increasing oscillation frequency show that the
solver remains robust as the sweeping speed and the temporal stiffness of the
forcing increase.

Overall, the moving-geometry results draw a coherent picture: the proposed
space-time cut-cell formulation (i) preserves constant states through a
discrete geometric conservation law, (ii) maintains global conservation across
topology changes within a time step, (iii) remains robust on severely
under-resolved moving geometries and (iv) achieves super-linear convergence
in practical moving-interface benchmarks in 2D and 3D, including multiple-body
motion and two-phase coupling. These properties provide the essential numerical
building blocks for more complex sharp-interface multiphysics.

The most natural next step is to move from prescribed motion to free-boundary
dynamics, where interface velocity is determined by physics rather than imposed.
The present space-time machinery is directly aligned with Stefan-type phase-change
models: the method already provides conservative phase balances, accurate
approximations of normal fluxes on moving interfaces and a robust handling of
discrete events as small phase volumes appear or vanish. Beyond pure diffusion,
extending the framework to advection-diffusion and to incompressible two-phase
Navier-Stokes with sharp interfaces would enable fully coupled problems with
density/viscosity jumps, interfacial stress continuity and thermal/mass transfer.
In that setting, the time-integrated geometric operators developed here can be
reused for diffusive and elliptic (pressure) terms, while additional space-time
consistent cut-cell operators are required for convection and for interface-consistent
momentum coupling. On the numerical side, an important direction is to strengthen the coupling
between interface advancement and the bulk PDE solves. In the present study, the
geometry is prescribed but in free-boundary settings the interface location and
the interfacial fluxes must be updated consistently within each time step. This
calls for tighter interface-tracking strategies together with coupled or iterative
interface-PDE algorithms that enforce the jump conditions and the interface
kinetics (e.g.\ Stefan-type laws) at the discrete level. This will make the method
a practical tool for large-scale, interface-resolved free-boundary multiphysics.

\newpage
\appendix

\section{Summary table for Space-time quantities notations}
\label{app:tablest}

\begin{table}[h!]
\centering
\renewcommand{\arraystretch}{2.0}
\begin{tabular}{lll}
\hline
\textbf{Quantity} & \textbf{Definition using $\langle \cdot , \cdot \rangle$} & \textbf{Equivalent integral form} \\[3pt]
\hline
T-i cell volume &
$\displaystyle
\mathcal{V}_{n+\sfrac{1}{2},i,j}^{\pm}
= \left\langle ]t_{n},t_{n+1}[,\,V_{i,j}^{\pm}(t) \right\rangle
$ &
$\displaystyle \int_{t_{n}}^{t_{n+1}}
  \!\!\int_{\Omega_{i,j}^{\pm}(t)} 1\,\mathrm{d}V\,\mathrm{d}t$ \\[6pt]

T-i face area ($x$-normal) &
$\displaystyle
\mathcal{A}_{n+\sfrac{1}{2},i-\sfrac{1}{2},j}^{1\pm}
= \left\langle ]t_{n},t_{n+1}[,\,A_{i-\sfrac{1}{2},j}^{1\pm}(t) \right\rangle
$ &
$\displaystyle \int_{t_{n}}^{t_{n+1}}
  \!\!\int_{\Sigma_{j}^{1\pm}(t,x_{i-\sfrac{1}{2}})} 1\,\mathrm{d}S\,\mathrm{d}t$ \\[6pt]

T-i face area ($y$-normal) &
$\displaystyle
\mathcal{A}_{n+\sfrac{1}{2},i,j-\sfrac{1}{2}}^{2\pm}
= \left\langle ]t_{n},t_{n+1}[,\,A_{i,j-\sfrac{1}{2}}^{2\pm}(t) \right\rangle
$ &
$\displaystyle \int_{t_{n}}^{t_{n+1}}
  \!\!\int_{\Sigma_{i}^{2\pm}(t,y_{j-\sfrac{1}{2}})} 1\,\mathrm{d}S\,\mathrm{d}t$ \\[6pt]

T-a cell centroid ($x$-coord.) &
$\displaystyle
\mathcal{X}_{n+\sfrac{1}{2},i,j}^{\pm}
= 
\frac{\left\langle ]t_{n},t_{n+1}[,\,\langle \Omega_{i,j}^{\pm}(t),x\rangle \right\rangle}
     {\mathcal{V}_{n+\sfrac{1}{2},i,j}^{\pm}}
$ &
$\displaystyle
\frac{\int_{t_{n}}^{t_{n+1}}\!\!\int_{\Omega_{i,j}^{\pm}(t)} x\,\mathrm{d}V\,\mathrm{d}t}
     {\int_{t_{n}}^{t_{n+1}}\!\!\int_{\Omega_{i,j}^{\pm}(t)} \mathrm{d}V\,\mathrm{d}t}$ \\[6pt]

T-a cell centroid ($y$-coord.) &
$\displaystyle
\mathcal{Y}_{n+\sfrac{1}{2},i,j}^{\pm}
= 
\frac{\left\langle ]t_{n},t_{n+1}[,\,\langle \Omega_{i,j}^{\pm}(t),y\rangle \right\rangle}
     {\mathcal{V}_{n+\sfrac{1}{2},i,j}^{\pm}}
$ &
$\displaystyle
\frac{\int_{t_{n}}^{t_{n+1}}\!\!\int_{\Omega_{i,j}^{\pm}(t)} y\,\mathrm{d}V\,\mathrm{d}t}
     {\int_{t_{n}}^{t_{n+1}}\!\!\int_{\Omega_{i,j}^{\pm}(t)} \mathrm{d}V\,\mathrm{d}t}$ \\[6pt]

T-i centroidal face area ($x$-dir.) &
$\displaystyle
\mathcal{B}_{n+\sfrac{1}{2},i,j}^{1\pm}
=\left\langle ]t_{n},t_{n+1}[,\,\langle
  \Sigma_{j}^{1\pm}(t,\mathcal{X}_{n+\sfrac{1}{2},i,j}^{\pm}),1\rangle \right\rangle
$ &
$\displaystyle
\int_{t_{n}}^{t_{n+1}}\!\!\int_{\Sigma_{j}^{1\pm}(t,\mathcal{X}_{n+\sfrac{1}{2},i,j}^{\pm})}
1\,\mathrm{d}S\,\mathrm{d}t$ \\[6pt]

T-i centroidal face area ($y$-dir) &
$\displaystyle
\mathcal{B}_{n+\sfrac{1}{2},i,j}^{2\pm}
=\left\langle ]t_{n},t_{n+1}[,\,\langle
  \Sigma_{i}^{2\pm}(t,\mathcal{Y}_{n+\sfrac{1}{2},i,j}^{\pm}),1\rangle \right\rangle
$ &
$\displaystyle
\int_{t_{n}}^{t_{n+1}}\!\!\int_{\Sigma_{i}^{2\pm}(t,\mathcal{Y}_{n+\sfrac{1}{2},i,j}^{\pm})}
1\,\mathrm{d}S\,\mathrm{d}t$ \\[6pt]

T-i staggered volume ($x$-dir.) &
$\tiny\begin{aligned}
\mathcal{W}_{n+\sfrac{1}{2},i-\sfrac{1}{2},j}^{1\pm}
&=\left\langle ]t_{n},t_{n+1}[,\,\left\langle
  (\!]\mathcal{X}_{n+\sfrac{1}{2},i-1,j}^{\pm},
  \mathcal{X}_{n+\sfrac{1}{2},i,j}^{\pm}\![\!
\right.\right.\\
&\qquad\left.\left.
  \times\!]
  y_{j-\sfrac{1}{2}},y_{j+\sfrac{1}{2}}\![)
  \cap \Omega^{\pm}(t),1
\right\rangle \right\rangle
\end{aligned}$ &

$\displaystyle
\int_{t_{n}}^{t_{n+1}}\!\!\int_{(\cdot)\cap\Omega^{\pm}(t)} 1\,\mathrm{d}V\,\mathrm{d}t$ \\[6pt]

T-i staggered volume ($y$-dir.) &
$\tiny\begin{aligned}
\mathcal{W}_{n+\sfrac{1}{2},i,j-\sfrac{1}{2}}^{2\pm}
&=\left\langle ]t_{n},t_{n+1}[,\,\left\langle
  (\!] x_{i-\sfrac{1}{2}},x_{i+\sfrac{1}{2}}\![\!
\right.\right.\\
&\qquad\left.\left.
  \times\!]
  \mathcal{Y}_{n+\sfrac{1}{2},i,j-1}^{\pm},
  \mathcal{Y}_{n+\sfrac{1}{2},i,j}^{\pm}\![)
  \cap \Omega^{\pm}(t),1
\right\rangle \right\rangle
\end{aligned}$ &

$\displaystyle
\int_{t_{n}}^{t_{n+1}}\!\!\int_{(\cdot)\cap\Omega^{\pm}(t)} 1\,\mathrm{d}V\,\mathrm{d}t$ \\[6pt]

\hline\hline

S-t averaged bulk variable &
$\displaystyle
\Tilde{\Phi}_{i,j}^{\omega\pm}
= 
\frac{\left\langle ]t_{n},t_{n+1}[,\,\langle \Omega_{i,j}^{\pm}(t),\phi^{\pm}(t)\rangle \right\rangle}
{\mathcal{V}_{n+\sfrac{1}{2},i,j}^{\pm}}
$ &
$\displaystyle
\frac{\int_{t_{n}}^{t_{n+1}}\!\!\int_{\Omega_{i,j}^{\pm}(t)} \phi^{\pm}(t)\,\mathrm{d}V\,\mathrm{d}t}
{\int_{t_{n}}^{t_{n+1}}\!\!\int_{\Omega_{i,j}^{\pm}(t)} \mathrm{d}V\,\mathrm{d}t}$ \\[6pt]

S-t averaged flux (x-dir.) &
$\displaystyle
\mathcal{Q}^{1\pm}_{\,n+\sfrac12,\,i+\sfrac12,\,j}
=
\frac{\left\langle ]t_{n},t_{n+1}[,\,
\langle \Sigma^{1\pm}_j(x_{i+\sfrac12}),\,q^{1\pm}\rangle \right\rangle}
{\mathcal{A}^{1\pm}_{\,n+\sfrac12,\,i+\sfrac12,\,j}}%,\quad \mathcal{A}^{1\pm}_{\,n+\sfrac12,\,i+\sfrac12,\,j}\neq 0
$ &
$\displaystyle
\frac{\int_{t_{n}}^{t_{n+1}}\!\!\int_{\Sigma^{1\pm}_j(t,x_{i+\sfrac12})} q^{1\pm}(t)\,\mathrm{d}S\,\mathrm{d}t}
{\int_{t_{n}}^{t_{n+1}}\!\!\int_{\Sigma^{1\pm}_j(t,x_{i+\sfrac12})} \mathrm{d}S\,\mathrm{d}t}$ \\[6pt]

S-t averaged source term &
$\displaystyle
\mathcal{R}_{\,n+\sfrac12,\,i,j}^{\pm}
=
\frac{\left\langle ]t_{n+1},t_{n+1}[,\,
\langle \Omega_{i,j}^{\pm}(t),\,r^{\pm}(t)\rangle \right\rangle}
{\mathcal{V}^{\pm}_{\,n+\sfrac12,\,i,j}}
$ &
$\displaystyle
\frac{\int_{t_{n+1}}^{t_{n+1}}\!\!\int_{\Omega_{i,j}^{\pm}(t)} r^{\pm}(t)\,\mathrm{d}V\,\mathrm{d}t}
{\int_{t_{n+1}}^{t_{n+1}}\!\!\int_{\Omega_{i,j}^{\pm}(t)} \mathrm{d}V\,\mathrm{d}t}$ \\[6pt]

S-t interface field average &
$\displaystyle \mathcal{F}_{i,j}(t) =
\frac{\left\langle ]t_{n+1},t_{n+1}[,\langle \Gamma_{i,j}(t), f(t) \rangle \right\rangle}{\int_{\Gamma_{i,j}(t)} 1 \, \mathrm{d}S \, \mathrm{d}t.}$ &
$\displaystyle \frac{\int_{t_{n}}^{t_{n+1}}\int_{\Gamma_{i,j}(t)} f(t)\,\mathrm{d}S \mathrm{d}t}
{\int_{t_{n}}^{t_{n+1}}\int_{\Gamma_{i,j}(t)} \mathrm{d}S\mathrm{d}t}$ \\[3pt]
\hline
\end{tabular}
\caption{Summary of space-time geometric quantities and semi-discrete averages expressed using nested integral notation $\langle ]t_{n},t_{n+1}[, \langle \Xi(t), f(t)\rangle \rangle$. T-i stands for "Time-integrated", T-a for "Time-averaged" and S-t for "Space-time"}
\label{tab:spacetime_geom_moments}
\end{table}

\bibliographystyle{plain}
\bibliography{references}

@article{glowinski_fictitious_1994,
	title = {A fictitious domain method for {Dirichlet} problem and applications},
	volume = {111},
	copyright = {https://www.elsevier.com/tdm/userlicense/1.0/},
	issn = {00457825},
	url = {https://linkinghub.elsevier.com/retrieve/pii/004578259490135X},
	doi = {10.1016/0045-7825(94)90135-X},
	language = {en},
	number = {3-4},
	urldate = {2025-12-23},
	journal = {Computer Methods in Applied Mechanics and Engineering},
	author = {Glowinski, Roland and Pan, Tsorng-Whay and Periaux, Jacques},
	month = jan,
	year = {1994},
	pages = {283--303},
}

@article{taira_immersed_2007,
	title = {The immersed boundary method: {A} projection approach},
	volume = {225},
	copyright = {https://www.elsevier.com/tdm/userlicense/1.0/},
	issn = {00219991},
	shorttitle = {The immersed boundary method},
	url = {https://linkinghub.elsevier.com/retrieve/pii/S0021999107001234},
	doi = {10.1016/j.jcp.2007.03.005},
	language = {en},
	number = {2},
	urldate = {2025-12-23},
	journal = {Journal of Computational Physics},
	author = {Taira, Kunihiko and Colonius, Tim},
	month = aug,
	year = {2007},
	pages = {2118--2137},
}

@article{peskin_flow_1972,
	title = {Flow patterns around heart valves: {A} numerical method},
	volume = {10},
	copyright = {https://www.elsevier.com/tdm/userlicense/1.0/},
	issn = {00219991},
	shorttitle = {Flow patterns around heart valves},
	url = {https://linkinghub.elsevier.com/retrieve/pii/0021999172900654},
	doi = {10.1016/0021-9991(72)90065-4},
	language = {en},
	number = {2},
	urldate = {2025-12-23},
	journal = {Journal of Computational Physics},
	author = {Peskin, Charles S},
	month = oct,
	year = {1972},
	pages = {252--271},
}

@article{mittal_immersed_2005,
	title = {{IMMERSED} {BOUNDARY} {METHODS}},
	volume = {37},
	issn = {0066-4189, 1545-4479},
	url = {https://www.annualreviews.org/doi/10.1146/annurev.fluid.37.061903.175743},
	doi = {10.1146/annurev.fluid.37.061903.175743},
	language = {en},
	number = {1},
	urldate = {2025-12-23},
	journal = {Annual Review of Fluid Mechanics},
	author = {Mittal, Rajat and Iaccarino, Gianluca},
	month = jan,
	year = {2005},
	pages = {239--261},
}

@article{mavriplis_unstructured_1995,
	title = {Unstructured mesh generation and adaptivity},
	url = {https://ntrs.nasa.gov/citations/19950020607},
	language = {en},
	urldate = {2025-12-23},
	author = {Mavriplis, D. J.},
	month = apr,
	year = {1995},
}

@article{hughes_isogeometric_2005,
	title = {Isogeometric analysis: {CAD}, finite elements, {NURBS}, exact geometry and mesh refinement},
	volume = {194},
	copyright = {https://www.elsevier.com/tdm/userlicense/1.0/},
	issn = {00457825},
	shorttitle = {Isogeometric analysis},
	url = {https://linkinghub.elsevier.com/retrieve/pii/S0045782504005171},
	doi = {10.1016/j.cma.2004.10.008},
	language = {en},
	number = {39-41},
	urldate = {2025-12-23},
	journal = {Computer Methods in Applied Mechanics and Engineering},
	author = {Hughes, T.J.R. and Cottrell, J.A. and Bazilevs, Y.},
	month = oct,
	year = {2005},
	pages = {4135--4195},
}

@misc{libat_cartesian_2025,
	title = {A {Cartesian} {Cut}-{Cell} {Two}-{Fluid} {Method} for {Two}-{Phase} {Diffusion} {Problems}},
	url = {http://arxiv.org/abs/2512.19407},
	doi = {10.48550/arXiv.2512.19407},
	urldate = {2025-12-23},
	publisher = {arXiv},
	author = {Libat, Louis and Selçuk, Can and Chénier, Eric and Chenadec, Vincent Le},
	month = dec,
	year = {2025},
	note = {arXiv:2512.19407},
}

@article{mccorquodale_cartesian_2001,
	title = {A {Cartesian} {Grid} {Embedded} {Boundary} {Method} for the {Heat} {Equation} on {Irregular} {Domains}},
	volume = {173},
	copyright = {https://www.elsevier.com/tdm/userlicense/1.0/},
	issn = {00219991},
	url = {https://linkinghub.elsevier.com/retrieve/pii/S0021999101969001},
	doi = {10.1006/jcph.2001.6900},
	language = {en},
	number = {2},
	urldate = {2025-12-19},
	journal = {Journal of Computational Physics},
	author = {McCorquodale, Peter and Colella, Phillip and Johansen, Hans},
	month = nov,
	year = {2001},
	pages = {620--635},
}

@article{johansen_cartesian_1998,
	title = {A {Cartesian} {Grid} {Embedded} {Boundary} {Method} for {Poisson}'s {Equation} on {Irregular} {Domains}},
	volume = {147},
	copyright = {https://www.elsevier.com/tdm/userlicense/1.0/},
	issn = {00219991},
	url = {https://linkinghub.elsevier.com/retrieve/pii/S0021999198959654},
	doi = {10.1006/jcph.1998.5965},
	language = {en},
	number = {1},
	urldate = {2025-06-26},
	journal = {Journal of Computational Physics},
	author = {Johansen, Hans and Colella, Phillip},
	month = nov,
	year = {1998},
	pages = {60--85},
}

@article{schwartz_cartesian_2006,
	title = {A {Cartesian} grid embedded boundary method for the heat equation and {Poisson}’s equation in three dimensions},
	volume = {211},
	copyright = {https://www.elsevier.com/tdm/userlicense/1.0/},
	issn = {00219991},
	url = {https://linkinghub.elsevier.com/retrieve/pii/S002199910500286X},
	doi = {10.1016/j.jcp.2005.06.010},
	language = {en},
	number = {2},
	urldate = {2025-11-13},
	journal = {Journal of Computational Physics},
	author = {Schwartz, Peter and Barad, Michael and Colella, Phillip and Ligocki, Terry},
	month = jan,
	year = {2006},
	pages = {531--550},
}

@article{farhat_discrete_2001,
	title = {The {Discrete} {Geometric} {Conservation} {Law} and the {Nonlinear} {Stability} of {ALE} {Schemes} for the {Solution} of {Flow} {Problems} on {Moving} {Grids}},
	volume = {174},
	issn = {0021-9991},
	url = {https://www.sciencedirect.com/science/article/pii/S0021999101969323},
	doi = {10.1006/jcph.2001.6932},
	number = {2},
	urldate = {2025-11-04},
	journal = {Journal of Computational Physics},
	author = {Farhat, C. and Geuzaine, P. and Grandmont, C.},
	month = dec,
	year = {2001},
	note = {Farhat2001},
	pages = {669--694},
}

@article{tarzia_bibliography_2000,
	title = {A bibliography on moving-free boundary problems for the heat-diffusion equation. {The} stefan and related problems},
	volume = {2},
	issn = {15154904},
	url = {http://web.austral.edu.ar/descargas/facultad-cienciasEmpresariales/mat/Tarzia-MAT-SerieA-2(2000).pdf},
	doi = {10.26422/MAT.A.2000.2.tar},
	urldate = {2025-07-08},
	journal = {MAT Serie A},
	author = {Tarzia, Domingo Alberto},
	month = jul,
	year = {2000},
	pages = {1--297},
}

@article{gabbard_high-order_2024,
	title = {A high-order finite difference method for moving immersed domain boundaries and material interfaces},
	volume = {507},
	issn = {00219991},
	url = {https://linkinghub.elsevier.com/retrieve/pii/S0021999124002286},
	doi = {10.1016/j.jcp.2024.112979},
	language = {en},
	urldate = {2025-07-04},
	journal = {Journal of Computational Physics},
	author = {Gabbard, James and Van Rees, Wim M.},
	month = jun,
	year = {2024},
	pages = {112979},
}

@article{gibou_second-order-accurate_2002,
	title = {A {Second}-{Order}-{Accurate} {Symmetric} {Discretization} of the {Poisson} {Equation} on {Irregular} {Domains}},
	volume = {176},
	copyright = {https://www.elsevier.com/tdm/userlicense/1.0/},
	issn = {00219991},
	url = {https://linkinghub.elsevier.com/retrieve/pii/S0021999101969773},
	doi = {10.1006/jcph.2001.6977},
	language = {en},
	number = {1},
	urldate = {2025-07-04},
	journal = {Journal of Computational Physics},
	author = {Gibou, Frederic and Fedkiw, Ronald P. and Cheng, Li-Tien and Kang, Myungjoo},
	month = feb,
	year = {2002},
	pages = {205--227},
}

@book{alexiades_mathematical_2018,
	edition = {1},
	title = {Mathematical {Modeling} of {Melting} and {Freezing} {Processes}},
	isbn = {9780203749449},
	url = {https://www.taylorfrancis.com/books/9781351433280},
	language = {en},
	urldate = {2025-06-30},
	publisher = {Routledge},
	author = {Alexiades, Vasilios and Solomon, Alan D.},
	month = may,
	year = {2018},
	doi = {10.1201/9780203749449},
}

@article{mccorquodale_high-order_2011,
	title = {A high-order finite-volume method for conservation laws on locally refined grids},
	volume = {6},
	issn = {2157-5452, 1559-3940},
	url = {http://msp.org/camcos/2011/6-1/p01.xhtml},
	doi = {10.2140/camcos.2011.6.1},
	language = {en},
	number = {1},
	urldate = {2025-06-26},
	journal = {Communications in Applied Mathematics and Computational Science},
	author = {McCorquodale, Peter and Colella, Phillip},
	month = mar,
	year = {2011},
	pages = {1--25},
}

@article{calhoun_cartesian_2000,
	title = {A {Cartesian} {Grid} {Finite}-{Volume} {Method} for the {Advection}-{Diffusion} {Equation} in {Irregular} {Geometries}},
	volume = {157},
	copyright = {https://www.elsevier.com/tdm/userlicense/1.0/},
	issn = {00219991},
	url = {https://linkinghub.elsevier.com/retrieve/pii/S0021999199963696},
	doi = {10.1006/jcph.1999.6369},
	language = {en},
	number = {1},
	urldate = {2025-06-26},
	journal = {Journal of Computational Physics},
	author = {Calhoun, Donna and LeVeque, Randall J.},
	month = jan,
	year = {2000},
	pages = {143--180},
}

@book{bird_transport_2007,
	address = {New York},
	edition = {Revised ed},
	title = {Transport phenomena},
	isbn = {9780470115398},
	language = {eng},
	publisher = {Wiley},
	author = {Bird, Robert Byron and Stewart, Warren E. and Lightfoot, Edwin N.},
	year = {2007},
}

@article{chierici_optimized_2022,
	title = {An optimized {Vofi} library to initialize the volume fraction field},
	volume = {281},
	issn = {00104655},
	url = {https://linkinghub.elsevier.com/retrieve/pii/S0010465522002259},
	doi = {10.1016/j.cpc.2022.108506},
	language = {en},
	urldate = {2025-05-20},
	journal = {Computer Physics Communications},
	author = {Chierici, A. and Chirco, L. and Le Chenadec, V. and Scardovelli, R. and Yecko, Ph. and Zaleski, S.},
	month = dec,
	year = {2022},
	pages = {108506},
}

@article{stefan_ueber_1891,
	title = {Ueber die {Theorie} der {Eisbildung}, insbesondere über die {Eisbildung} im {Polarmeere}},
	volume = {278},
	issn = {0003-3804, 1521-3889},
	url = {https://onlinelibrary.wiley.com/doi/10.1002/andp.18912780206},
	doi = {10.1002/andp.18912780206},
	language = {en},
	number = {2},
	urldate = {2025-05-16},
	journal = {Annalen der Physik},
	author = {Stefan, J.},
	month = jan,
	year = {1891},
	pages = {269--286},
}

@article{ho_discrete_2021,
	title = {Discrete embedded boundary method with smooth dependence on the evolution of a fluid‐structure interface},
	volume = {122},
	issn = {0029-5981, 1097-0207},
	url = {https://onlinelibrary.wiley.com/doi/10.1002/nme.6455},
	doi = {10.1002/nme.6455},
	language = {en},
	number = {19},
	urldate = {2025-05-16},
	journal = {International Journal for Numerical Methods in Engineering},
	author = {Ho, Jonathan and Farhat, Charbel},
	month = oct,
	year = {2021},
	pages = {5353--5383},
}

\end{document}